\documentclass[11pt]{article}
\usepackage{emulateapj,epsf,apjfonts}

\def\msun{{\rm\,M_\odot}}
\def\msun{{\rm\,M_\odot}}
\def\kpc{{\rm\,kpc}}
\def\kms{{\rm\,km/s}}

\def\gyr{{\rm\,Gyr}}
\def\pc{{\rm\,pc}}
\def\spose#1{\hbox to 0pt{#1\hss}}
\def\lta{\mathrel{\spose{\lower 3pt\hbox{$\mathchar"218$}}
     \raise 2.0pt\hbox{$\mathchar"13C$}}}
\def\gta{\mathrel{\spose{\lower 3pt\hbox{$\mathchar"218$}}
     \raise 2.0pt\hbox{$\mathchar"13E$}}}

\def\sech{\mathop{\rm sech}\nolimits}
\begin{document}

\title{Effect of the Milky Way on Magellanic Cloud structure}

\author{Martin D. Weinberg}
\affil{Department of Physics \& Astronomy, University of
  Massachusetts, Amherst, MA 01003-4525, USA}

\begin{abstract}
  A combination of analytic models and n-body simulations implies that
  the structural evolution of the Large Magellanic Cloud (LMC) is
  dominated by its dynamical interaction with the Milky Way.  Although
  expected at some level, the scope of the involvement has significant
  observational consequences.  First, LMC disk orbits are torqued out
  of the disk plane, thickening the disk and populating a spheroid.
  The torque results from direct forcing by the Milky Way tide and,
  indirectly, from the drag between the LMC disk and its halo
  resulting from the induced precession of the LMC disk.  The latter
  is a newly reported mechanism that can affect all satellite
  interations.  However, the overall torque can not isotropize the
  stellar orbits and their kinematics remains disk-like.  Such a
  kinematic signature is observed for nearly all LMC populations.  The
  extended disk distribution is predicted to increase the microlensing
  toward the LMC.  Second, the disk's binding energy slowly decreases
  during this process, puffing up and priming the outer regions for
  subsequent tidal stripping.  Because the tidally stripped debris
  will be spatially extended, the distribution of stripped stars is
  much more extended than the H{\sc I} Magellanic Stream.  This is
  consistent with upper limits to stellar densities in the gas stream
  and suggests a different strategy for detecting the stripped stars.
  And, finally, the mass loss over several LMC orbits is predicted by
  n-body simulation and the debris extends to tens of kiloparsecs from
  the tidal boundary.  Although the overall space density of the
  stripped stars is low, possible existence of such intervening
  populations have been recently reported and may be detectable using
  2MASS.
\end{abstract}

\section{Introduction}

The Magellanic Clouds are a natural laboratory for investigating the
evolution of stellar populations in dynamically interacting systems.
Their populations are well studied and provide a basis for standard
candle and population evolution studies.  In addition, a variety of
dynamical studies and simulations exploit the Milky
Way--Clouds--Magellanic Stream interaction to both infer its complex
history and constrain Milky Way mass models (Fujimoto \& Sofue 1976,
Lin \& Lynden-Bell 1977, Davies \& Wright 1977, Lin \& Lynden-Bell
1982, Murai \& Fujimoto 1986, Lin et al. 1995; see Lin et al. for a
thorough historical discussion).  The link between chemical
evolution and externally driven structural evolution is a hard
problem, and to date, there has been little work devoted to a
self-consistent dynamical picture of Cloud evolution.  To this end,
this paper focuses on one specific aspect---the dynamical interaction
between the Milky Way and the Large Magellanic Cloud---and ignores the
likelihood of a significant interaction with the Small Magellanic
Cloud (SMC) in the past or the possibility that the SMC originated in
a tidal disruption event.  This simplified scenario by itself admits a
rich set of interacting mechanisms.

Recent work by myself and others points out that time-dependent tidal
forcing can have significant evolutionary consequences for globular
clusters and dwarf or cannibalized galaxies (Chernoff et al.  1986,
Aguilar et al. 1988, Weinberg 1994c, Gnedin \& Ostriker 1997, Murali
\& Weinberg 1997ab, Vesperini 1997, Weinberg
1997).  The same
physics applies to non-isotropic distributions such as disks or disks
embedded in halos.  For example, Sellwood et al.
(1998) explored the importance of these
resonant mechanisms to thickening host disks by dwarfs and excitation
of bending waves.  Weinberg (1998, Paper I) found
that the resonant interaction between the Milky Way and LMC is
sufficient to excite a warp and cause lopsided asymmetries, depending
on the Galactic halo potential and LMC mass.  In that work, the LMC
was structurally fixed.  Here, we turn the tables by structurally
fixing the Milky Way and applying the same physics to LMC evolution.

It is straightforward to see that the magnitude of such a disturbance
to the LMC is large.  Either the subtended size of the LMC on the sky
or the rotation curve combined with an estimate of the Milky Way mass
enclosed in the LMC orbit leads to a tidal radius of approximately 11
kpc.  At 5 kpc from the center, the ratio of the tidal force to the
self force has only dropped to approximately 20\% assuming a flat
rotation curve, a significant perturbation.  Although this ratio drops
quickly further inward, the effect of the tidal force is amplified by
a spectrum of resonances between the LMC--Milky Way orbital
frequencies and internal LMC orbital frequencies.  Simultaneously, the
LMC disk axis precesses due to the coupling with its orbit about the
Galaxy.  This induces an additional interaction between the LMC disk
and halo.  Altogether, these mechanisms result in enhanced angular
momentum and energy transfer between the orbit and internal motions.
They thicken the disk, populate the spheroid and drive mass loss.  The
latter mechanism has direct analogy to globular cluster evolution (see
refs. cited above).

This paper explores this basic picture as follows.  First,
\S\ref{sec:massmodel} summarizes the inference of the LMC orbit and
mass needed to estimate the time-dependent tidal force.  We will then
explore the underlying dynamical interaction in several steps.
Analytically, effectively irreversible changes in energy and angular
momentum occur at resonances between the frequencies of the applied
perturbation and the stellar orbital frequencies.  First, I will
describe the results of a restricted computation which sums the effect
of all the resonances directly.  This idealized model treats the
evolution of a disk without self-gravity in a fixed halo potential
(\S\ref{sec:boltz}).  We find that the disk is notably heated and
thickened in several gigayears.  Although the omission of self-gravity
surely leads to an overestimate of the heating, the simple model
serves to illustrate the potential importance of the underlying
physical mechanisms.

This example is followed up in \S\ref{sec:nbody} with a full n-body
simulation using the force algorithm described in Weinberg
(1999).  This code uses a basis expansion
tailored to the density profile and is well-suited to following a
slowly changing system.  The collisionless evolution is
gravitationally self-consistent with the caveat that the Galactic mass
model and LMC orbit remains fixed.  The n-body results substantiate
the simple restricted example in \S\ref{sec:boltz}, although the rate
of disk thickening is smaller due to the self-gravity of the disk.
Specifically, the simulations predict a thickening rate of 70 pc/Gyr
at a roughly constant rate over the duration (about 4 Gyr).  The tail
of the torqued distribution populates the LMC halo region.  At the
same time, the energy input does work on the potential and causes
overall expansion of the disk.  This offsets the increase in velocity
dispersion that might be observed from heating in a fixed potential;
in fact, expansion wins and the velocity dispersion observed at a
$45^\circ$ inclination at one disk scale length is very slowly
dropping.

These results lead to a number of interesting predictions and
implications.  First, the stellar component should be as extended as
the halo.  Moreover, this is done without isotropizing the
distribution since a modest change in direction of the orbital plane
(and therefore its angular momentum vector) is relatively easy.  In
fact, Olszewski et al. (1996) outline
the evidence that nearly all components of the LMC have disk-like
kinematics regardless of their extent.  Second, the dark halo and the
kinematically evolved and extended stellar component are
preferentially stripped.  The stripped material continues to orbit
with the LMC and slowly spread in phase.  Because the stripped stars
are not part of the disk, they do not lie along the H{\sc I}-defined
Magellanic Stream but rather in a much more diffuse distribution.  In
other words, this scenario does not suggest looking in the gas stream
for the stars.  Finally, a thickened bound stellar component in the
LMC and an extended unbound cloud surrounding the LMC will increase
the rate of self microlensing and we will estimate the effect in
\S\ref{sec:mulens}.  A final discussion with implications for the LMC
and satellite systems in general is presented in
\S\ref{sec:discussion} followed by a summary of results in
\S\ref{sec:summary}.

\section{Mass, Structure and Orbit of LMC}
\label{sec:massmodel}

There have been a wide variety of LMC censuses, most of which treat
the LMC as a galaxy and use the standard mass and mass density
estimates: rotation curves, star counts, surface brightness profiles.
Two relatively recent rotation curve studies, Meatheringham et al.
(1988) and Schommer et al.
(1992), estimate LMC masses of
$6\times10^9\msun$ and $1.5\times10^{10}\msun$.  Similar limits follow
from carbon star studies by Kunkel et al.
(1997b). The main difference between these
determinations is not the value of $V_c$ for the Cloud but the radial
extent of the rotation curve.  Alternatively, from the Milky Way's
point of view, the LMC is similar to an oversized globular cluster.
Its tidal radius is measurable and depends on both the Milky Way
rotation curve and the LMC mass (and, weakly, its profile).  A
preliminary estimate of the tidally inferred LMC mass (Nikolaev \&
Weinberg 1998) yields
$2\times10^{10}\msun$ but is consistent with the Schommer et al.
estimate.  A brief description of this result is provided in the
Appendix.

Recent estimates of the LMC space velocity from archival plate (Jones
et al. 1994) and Hipparcos (Kroupa \&
Bastian 1997) proper motions both lead to
consistent estimates of the LMC orbital plane.  The procedure used
here to estimate the orbit is described in Weinberg
(1995).  For the Milky Way halo, I choose a
$W_0=3$ King model (1966) with $r_t=200\kpc$ and mass
scaled to $4\times10^{11}\msun$ (Kochanek 1996).
The rotation curve due to the Galaxy, then, is approximately flat in
the region of the LMC orbit.  Together with a disk, the overall
rotation curve is a plausible representation of the observed Milky
Way.  With this halo model, both the space velocities estimated by
Jones et al.  and Kroupa \& Bastian yield a similar perigalacticon of
46 kpc with apogalacticons of 72 kpc and 120 kpc.  For lack of
motivation to favor one of these over the other, I adopt the mean
apogalacticon.

\section{Milky Way heating of the LMC}
\label{sec:heat}

Paper I described the excitation of structure in the Milky Way halo
due to non-local resonant excitation that occurs well inside the LMC's
orbit.  From the LMC's point of view, the Milky Way is in orbit about
the LMC and the same dynamical coupling that raises wakes in the halo
affects the LMC disk.  This periodic forcing changes the angular
momenta of orbits at commensurate frequencies and adds energy to the
disk.  In the absence of commensurabilities, the tidal forcing would
be adiabatically reversible and not lead to long-term evolution (e.g.
Weinberg 1994a). This is the same physics that
``heats'' stellar orbits in globular clusters (Weinberg 1994c, Murali
\& Weinberg 1997a, Gnedin \& Ostriker 1997).  In the globular cluster
case, however, one is primarily concerned with the work done by the
external perturbation.  In a disk, a change only in the orbital
angular momentum vector, with little energy transfer, can change the
disk morphology.

The first subsection below illustrates the evolution based on this
dynamical mechanism.  The calculation is straightforward in the
idealized scenario of a spheroid-dominated potential and axisymmetry
and without explicit disk self gravity.  It predicts significant
evolution on a gigayear time scale.  Including disk self gravity will
lengthen the time scale but nonetheless this relatively short time
scale motivates the more complete n-body treatment in
\S\ref{sec:nbody}.

\subsection{Solution of Boltzmann equation}
\label{sec:boltz}

\begin{figure*}
  \mbox{\epsfxsize=5.5in\epsfbox{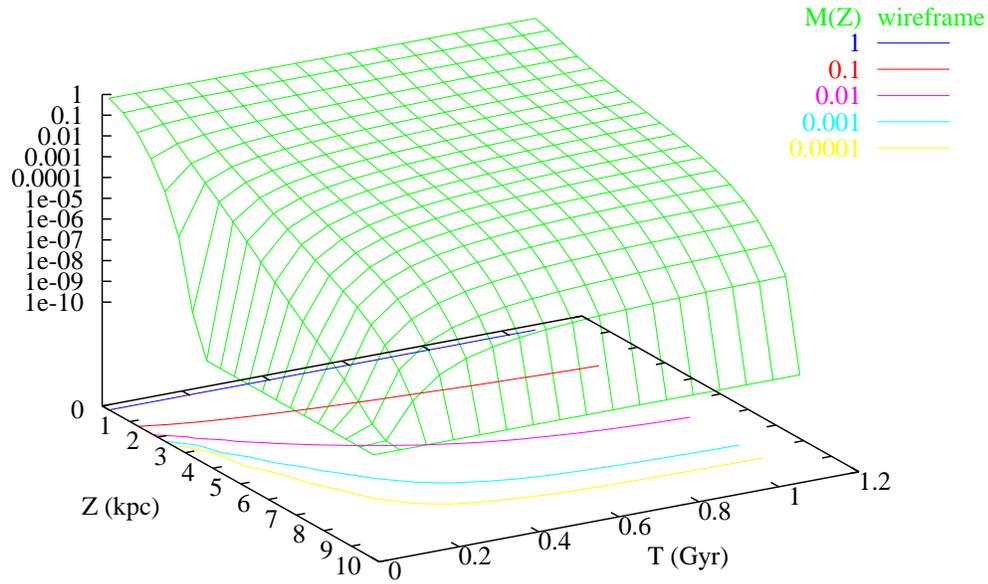}}
  \caption{LMC disk heating by the Milky Way.  Contours and wire frame
    show the cumulative distribution of stars at a height $Z$ or
    larger.  The curves show mass fractions $1, 10^{-1}, 10^{-2},
    10^{-3}, 10^{-4}$ from bottom to top.}
  \label{fig:lmcdisk1}
\end{figure*}

\begin{figure*}
  \mbox{\epsfxsize=5.0in\epsfbox{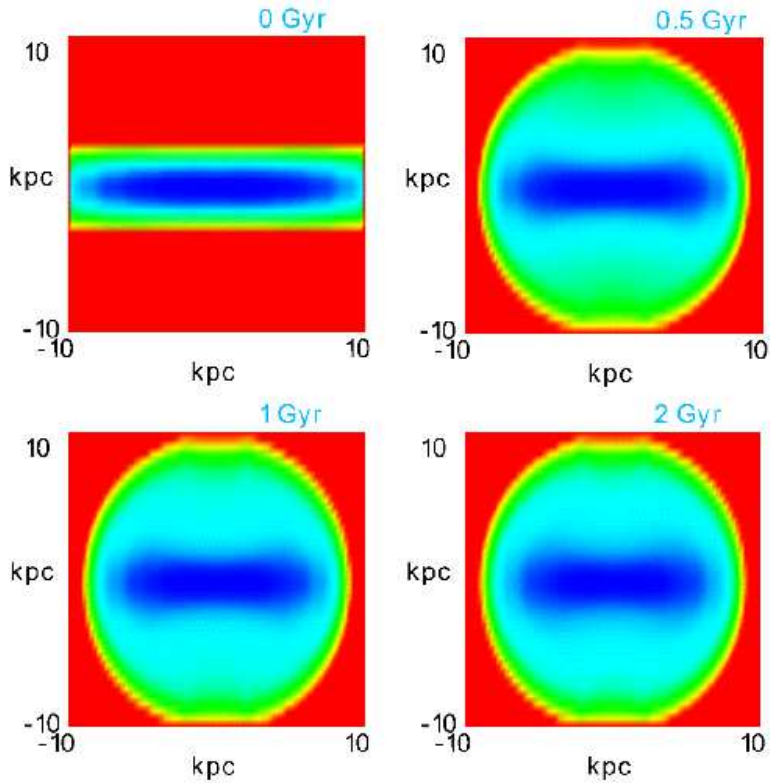}}
  \caption{The projected surface density distribution for
    the edge on view of disk at four times shown in Fig.
    \protect{\ref{fig:lmcdisk1}}.  The smooth color variation from red
    to green to blue reflects logarithmically change in projected
    surface density over six orders of magnitude from the peak.}
  \label{fig:lmcdisk2}
\end{figure*}

\begin{figure*}
  \mbox{\epsfxsize=6.5in\epsfbox{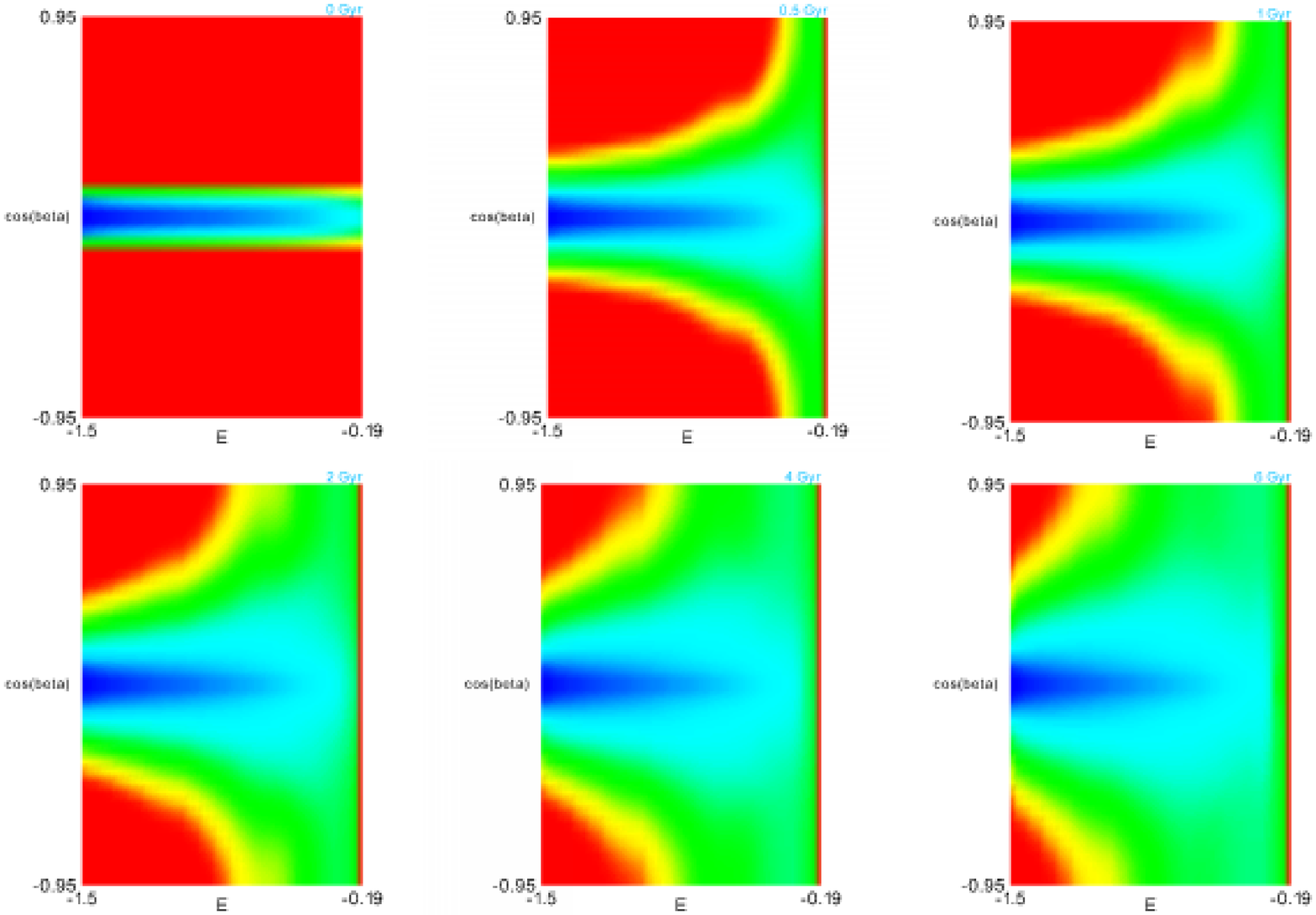}}
  \caption{As in Fig. \protect{\ref{fig:lmcdisk2}} but showing the
    phase space density in the $E$ and $\cos\beta\equiv J_z/J$
    plane.  NB: the evolution is both axisymmetric and symmetric
    around the midplane ($\cos\beta=0$).  }
  \label{fig:lmcdisk_eb}
\end{figure*}

To estimate the evolution, I present a solution of the time-dependent
collisional Boltzmann equation for orbits in a fixed potential. The
angular momenta of individual stars change during passage through
resonances as the disk slowly evolves. The change in the conserved
angular momenta depends on the direction that an orbit crosses a
particular resonance (see Henrard 1982 for
discussion).  A galaxy will have different
phase-space densities on either side of the resonance resulting in a
net gain or loss for the passage.  The net change in the phase-space
distribution function, then, due to the resonant heating takes the
form of a collisional Boltzmann equation where the right-hand-side
collision term depends on the gradient of the phase-space distribution
function (see Appendix for additional detail).  For simplicity, we
assume that the background gravitational potential is constant in
time, dominated by the halo.  The now linear partial differential
equation may be solved by finite-difference on a three-dimensional
grid (e.g.  $E$, $J$, $J_z$).  The z-axis is perpendicular to the disk
plane.  The ratio of the z-axis angular momentum to the total angular
momentum is the cosine of the orbital-plane inclination angle,
$\beta$: $\cos\beta = J_z/J$.  At every time step, the potential is
recomputed and any phase space whose stars have apocenters larger than
the tidal radius are deleted from the grid.  Although, these weakly
bound stars may linger near the tidal boundary for some time in
reality (Lee \& Ostriker 1987), this {\em one
  way} tidal boundary is easy to implement.  A $W_0=1.5$ King model
was chosen to represent the LMC gravitational potential and
approximately fits the rotation curve.

Figure \ref{fig:lmcdisk1} shows the cumulative distribution of mass
above the disk plane as the system evolves, $M(Z)$.  After
approximately 1 Gyr, 1\% of the disk mass has a height larger than 6
kpc and 10\% above 3 kpc.  The thickening occurs from the outside in,
appearing as a flared population that fills in at smaller radii with
time.  This leads to a very thick disk or flattened spheroid
population.

Figure \ref{fig:lmcdisk2} shows the edge-on projected surface mass
density.  One sees that the tidal envelope is filled in a gigayear,
and over longer time scales the disk scale height is increasing (cf.
the $10^{-1}$ contour in Fig. \ref{fig:lmcdisk1}).  This trend is more
apparent in phase space: the orbits at low binding energy are heated
first and those at successively higher binding energy as time goes on.
This is clearly seen in the energy-orbital inclination
($E$--$\cos\beta$) projection of the phase space distribution (Fig.
\ref{fig:lmcdisk_eb}): the high binding energy inclined orbits---the
upper and lower left corners--are successively filled in with time.

Also worthy of note is that $\log M(Z)$ is roughly linear with $Z$ at
times larger than 1 Gyr.  This suggests an exponential profile which
has been recently reported for the RR Lyrae distribution in the LMC
halo (Kinman et al.  1991).  The sharp
roll over at the tidal radius is suggestive of the observed star-count
profile but may be an artifact of the one-way tidal boundary.

\subsection{N-body solution}
\label{sec:nbody}

The idealized semi-analytic model suggests that the LMC disk structure
will change on a $10^9$ year time scale, roughly an LMC orbital time.
Without disk self gravity, the calculation in \S\ref{sec:boltz} is
expected to overestimate the disk thickening due to resonant heating,
although it may underestimate the thickening due to self-consistent
readjustment to the external work.  In this section, we examine the
details of the evolution by n-body simulation.  To limit the number of
parameters and difficulty of the simulation, we ignore orbital decay.

\subsubsection{Force solver}
\label{sec:method}

In order to estimate the evolution on large scales over long time
periods, we need to suppress small scale noise as much as possible and
this need is satisfied by the biorthogonal expansion technique (e.g.
Clutton-Brock 1972, 1973, Kalnajs 1976, Fridman \& Polyachenko 1984,
Hernquist \& Ostriker 1992).  The approach uses the eigenfunctions of the
Laplacian to construct a complete set of orthogonal functions that
satisfies the Poisson equation.  The projection of the ensemble
particle positions on each member of the orthogonal series yields a
set of coefficients, similar to determining a potential from a charge
distribution in electrostatics.  These coefficients may then be used
to describe the density, gravitational potential or force in the
expansion code.  The most efficient implementations use analytically
derived recursion relations to generate the functions.  The expansion
converges quickly and is most efficient when the lowest order function
is a good fit to the underlying equilibrium.  However, most
astronomical distributions do not match the available sets of special
functions.  The problem described in this paper motivated the
algorithm developed in Weinberg (1999) which
allows the adaptive construction of both spherical and
three-dimensional cylindrical bases and this algorithm is used here.

Another advantage of this force solver is its ability to separately
follow distinct kinematic components.  Each component may be tied to a
basis tailored to its geometry; this helps remove the bottleneck in
simultaneously resolving multiple spatial scales.  In particular, we
can assign the halo particles to a spherical basis and the disk
particles to a cylindrical basis.  These are gravitationally coupled
through the force evaluation.  This procedure is easily extended.  For
example, a simulation which follows the response of the Milky Way halo
simultaneously can be implemented by specifying an appropriate basis
for the halo and following the halo response and its back-reaction on
the LMC directly.  This would increase the run time of simulations
described here by only about 30\%.  These simulations will be
performed in the next phase of this project.  

Experimentation reveals that multiple expansion centers may introduce
numerical feedback and excite oscillations.  In principle, each
component may be uncoupled as long as feedback is suppressed.  For
simulations here, the halo expansion center is tied to the disk
center.  The force solver can easily resolve small offsets correctly
so this is not a limitation for this application.

A parallel code with these features is implemented on a Linux-based
cluster using MPI.  Each node in the cluster has two processors; the
algorithm was multi-threaded on each node to reduce the memory
overhead.  This allows the computation to be in core for all harmonic
orders used here.  The code performs load balancing but this is
usually not required for the dedicated cluster.  In practice, the
average CPU load efficiency is approximately 90\% for $N=10^5$ on 32
processors and improves for larger numbers of particles.

\subsubsection{Parameters}
\label{sec:params}

The resolution of the force computation is set both by controlling the
truncation in the series expansion and particle number.  In the
spherical case, one sets the maximum radial order, $n_{max}$ and the
maximum angular order $l_{max}$.  In the cylindrical case, to start
one has three indices corresponding the maximum radial order,
$n^\prime_{max}$, the maximum angular order $m_{max}$, and maximum
vertical order.  However, as described in Weinberg
(1999), one can only adaptively choose the
radial profile for the three-dimensional cylindrical set.  For a
particular spatial distribution and basis, the signal-to-noise ratio
decreases with increasing order.  We can empirically find new
orthogonal functions to best represent the underlying particle
distribution to minimize noise.  Here, this is done once based on the
initial conditions (see Weinberg 1996, 1999 for
details) and yields a two-dimensional indexed
set, $\mu_{max}, m_{max}$.  For problems here, we choose $l_{max},
m_{max}=2$ or $4$, $n_{max}, n^\prime_{nmax}=10$, and $\mu_{max}=10$.
Within these limits, low signal-to-noise ratio components may be
adaptively truncated.  The evolution shows little difference to tests
using larger series.  The vertical expansion also requires
specification of an outer boundary that is set to the initial tidal
radius.  Outside this radius, the force from the cylindrical component
reverts to a monopole spherical force estimate.  Finally, the time
step is chosen to be approximately $1/100$ of the shortest
characteristic time scale (inner disk vertical motion) and the time
evolution is followed using time-centered leap frog.

The total mass of the LMC is taken to be $2\times10^{10}\msun$,
divided evenly between a halo and an exponential-$\sech^2$ disk with
$a=1.6\kpc$ and $h=200\pc$ (e.g. Wu 1994).  The tidally
truncated halo is represented by a King model with a core radius small
enough to stabilize the disk against rapid bar formation.  The
potential of this halo profile includes a centrifugal potential during
disk generation to better account for the non-inertial forces in the
simulation.  The LMC halo and orbit and disk generation by the
quadratic programming technique is the same as that describe in
Weinberg (1998).  The simulations described here
used $N=4\times10^5$ particles with $1\times10^5$ in the disk and
$3\times10^5$ for the halo component.  The large number of halo
particles is needed to suppress the noise fluctuations which can
disturb the disk.  For the disk, the masses of the particles were
scaled to produce a uniform number density distribution in cylindrical
radius in order to better resolve the evolution of the outer disk.
The halo particles have equal mass.

The external force on the LMC is expressed in the non-inertial frame
of the center of mass of the LMC's orbit about the Milky Way:
\begin{eqnarray}
  {\bf f}_{tot}({\bf x}) &=& {\bf f}_{self}({\bf x}) + 
  {\bf F}_{gal}({\bf x}+{\bf R}(t)) -
  {\bf F}_{gal}({\bf R}(t)) \nonumber \\
  && -2{\bf\Omega}(t)\times{\bf v} -
  {\bf\Omega}(t)\times\left({\bf\Omega}(t)\times{\bf x}\right) -
  {\dot{\bf\Omega}}(t) \times {\bf x},
  \label{eq:tide}
\end{eqnarray}
where ${\bf x}$ is the position relative to the center of the LMC,
${\bf R}(t)$ is the center of the LMC relative to the Milky Way, ${\bf
  f}_{self}$ is the force of the LMC, ${\bf F}_{gal}$ is the force of
the Milky Way halo on the LMC and ${\bf\Omega}(t)$ is the
time-dependent azimuthal frequency of the LMC about its orbit.
Because the distance between the LMC tidal radius and the LMC
galactocentric radius is not small, the Galactic component tidal force
is evaluated by explicit difference rather than Taylor expansion of
the underlying halo potential model.

The LMC orbit, ${\bf R}(t)$, is assumed to be fixed, that is, we
ignore dynamical friction.  This was done to reduce the dynamical
complexity and number of parameters in favor of focusing on the
internal evolution.  Finally, because the main goal is to look at the
internal dynamical evolution of LMC, we ignore the effect of the SMC
(e.g.  Murai \& Fujimoto 1980, Lin et al.
1995).  Present day tidal
effects are dominated by the Milky Way, although a close encounter
with the SMC in past would also be a significant structure-changing
event for the LMC.

\subsubsection{Results}
\label{sec:results}

The galaxy is constructed to be in equilibrium in absence of the {\em
  time-dependent} tidal field.  During the first $6\times10^8$
years\footnote{Time units here are based on a circular velocity of
  $V_o=200\kms$ at the Solar circle and peak rotation curve velocity
  in the LMC of $V_{LMC}=75\kms$.}), the system achieves a new
approximate equilibrium.  A slow ramp-up of the tidal terms in
equation (\ref{eq:tide}) yield larger initial transients.  During the
initial virialization phase, the disk responds strongly in its outer
parts to the full non-inertial set of forces although the total mass
involved is small.  The inner disk oscillates as it phase mixes under
the fully consistent self-potential and external galactic potential.
Both effects have little effect on scale height.  A simulation with
the same initial conditions but with $n_{max}$ increased by a factor
of two shows the same behavior.

The LMC disk precesses under the torque from the Galaxy.  Uncorrected,
this would cause the expansion plane to drift away from the true disk
plane defined by the instantaneous mean angular momentum.  To follow
this, the disk bodies are ranked by binding energy and the lowest 2\%
are used to determine the disk angular momentum vector and expansion
center.  To damp any numerical feedback, both the expansion plane and
center are determined from a 100 time step running average.
    
\begin{figure*}[ht]
\mbox{
  \vbox to 3.5in { 
    \epsfxsize=3.0in\epsfbox{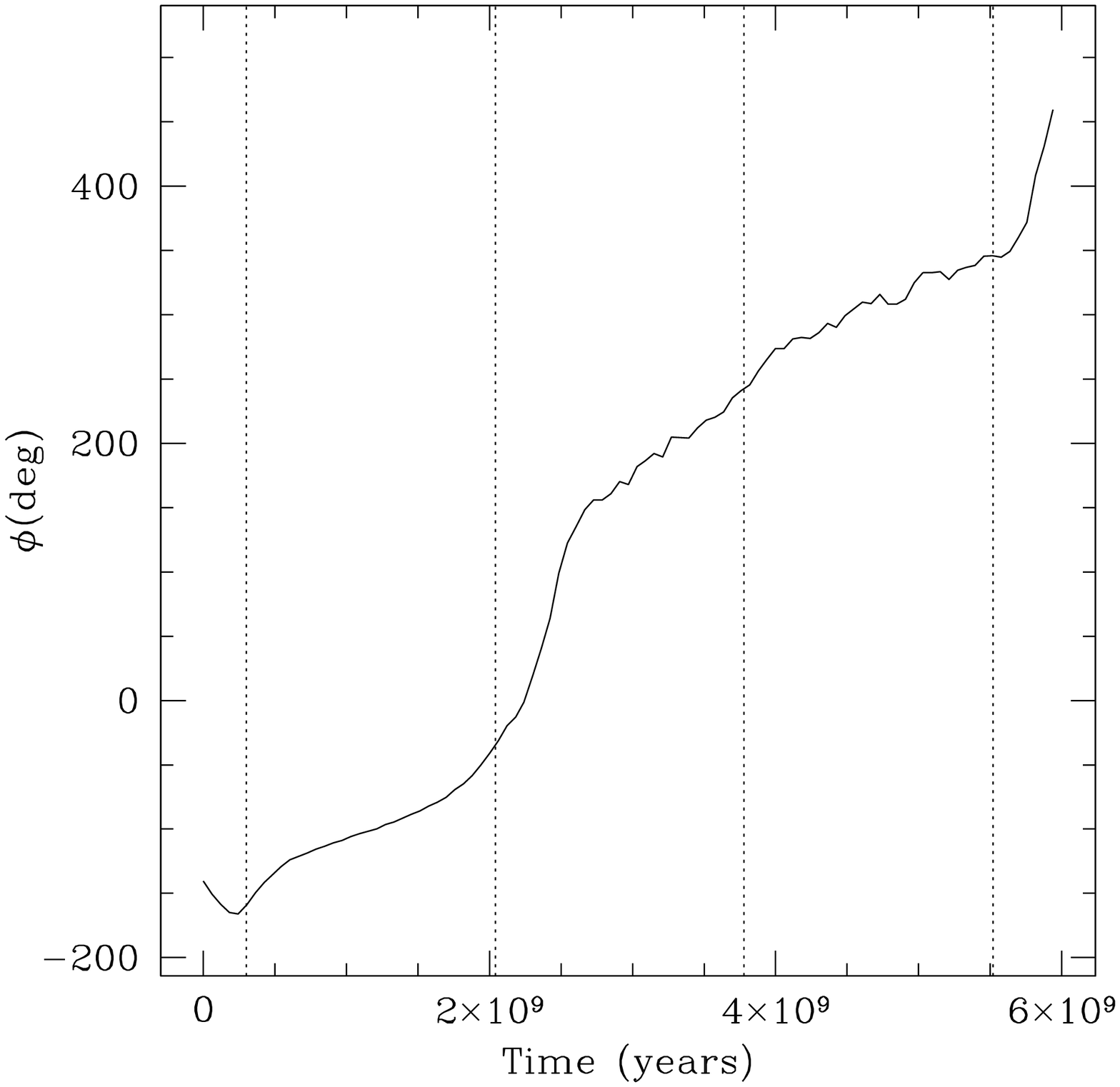}
    \hsize=3.0in
    \caption{Change in azimuth of the precessing disk with time.  The
      vertical dotted lines indicate perigalatica.
      \label{fig:azi}
      }
    }
  \hspace{0.5in}
  \vbox to 3.5in { 
    \epsfxsize=3.0in\epsfbox{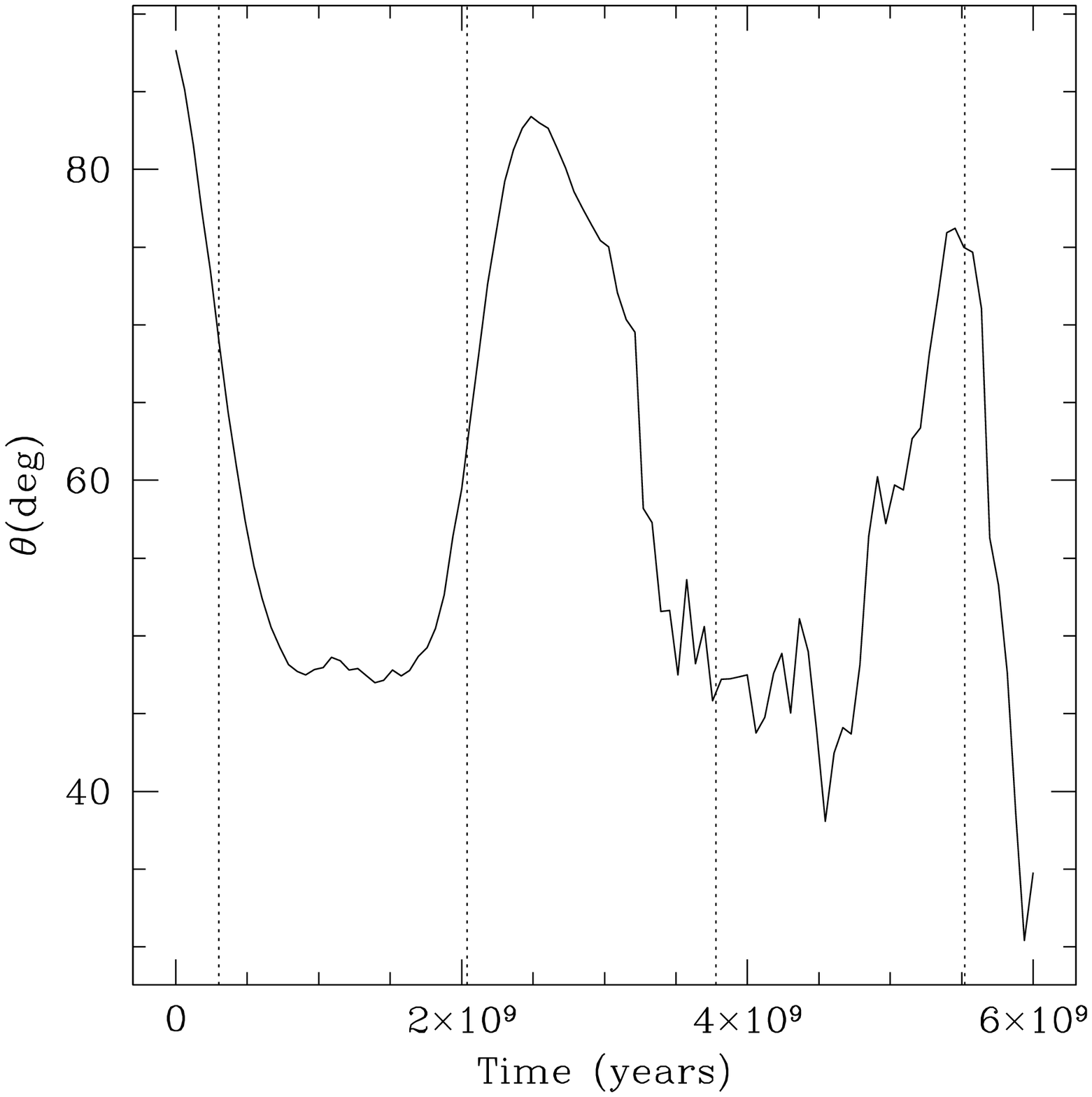}
    \hsize=3.0in
    \caption{
      Colatitude of angular momentum axis of precessing disk
      with time.  For comparison with Fig. \protect{\ref{fig:azi}}, the
      vertical dotted lines indicate perigalatica.
      \label{fig:elev}
      }
    }
  }
\end{figure*}

This precession is shown in Figure \ref{fig:azi} which shows the
azimuth of the LMC disk's angular momentum axis in the original frame.
The disk also nutates, as seen in Figure \ref{fig:elev}, because of
the initial transient.  The initial angle between the mean angular
momentum vector perpendicular to the disk and the Galactic center is
$45^\circ$.  Although a non-nutating system might be achieved by
iterating the initial angular momentum vector, there is little reason
to assume this is closer to the natural state, so no corrections have
been made.

\begin{figure*}[ht]
  \vbox{ 
    \epsfxsize=3.0in\epsfbox{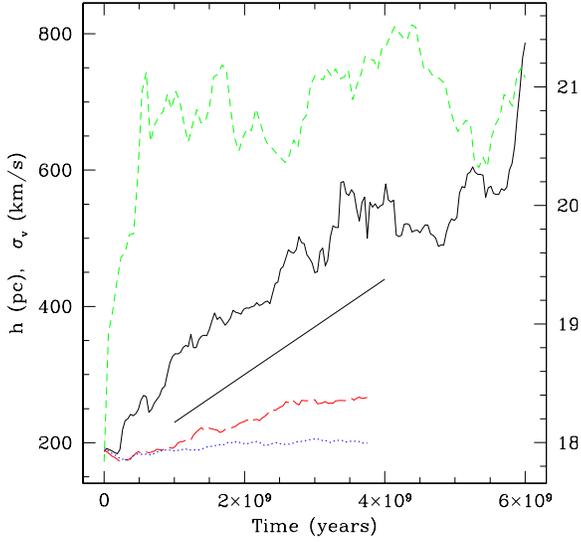}
    \caption{The
      scale height of the density distribution, \protect{$h$}
      (left-hand axis: solid, long dash, dotted) and the root variance
      of the velocity distribution, \protect{$\sigma_v$} (right-hand
      axis: upper dashed curve) for a line-of-sight inclined
      \protect{$45^\circ$} to the LMC disk.  The scale height of the
      LMC disk increases at a rate of 70 pc/Gyr (solid segment).  The
      dotted and long-dashed curves show the secular evolution in
      absence of the Galactic tidal field for $m=0$ terms only and all
      terms, respectively.  The velocity dispersion is nearly constant
      (note the small range in velocity) but decreases slowly on
      average after virialization.
    \label{fig:thick}
    }
  }
\end{figure*}

At the same time, the LMC halo and disk have a mutual self-gravitating
response that causes a slow $m=1$ oscillation of the disk density
center.  The existence of such weakly damped modes can be demonstrated
analytically using the methods described by Weinberg
(1994d).  Because of these modes and the fact
that mass and momentum loss in the Galactic tidal field is asymmetric,
the potential expansion is chosen to track the density center.
Although the force solver can handle this situation, the secular
heating of disk increases.  Tests without the tidal field confirm that
this effect does not dominate or obscure the tidal heating (cf. Fig.
\ref{fig:thick}).

The disk thickness is estimated from the density distribution in a
column through the LMC at a radius of a disk scale length.  The disk
plane is inferred by the same method used to orient the force
expansion (see \S\ref{sec:nbody}).  The line of sight inclination is
chosen to be $45^\circ$ and azimuthally averaged at one LMC disk scale
length.  The line of sight quantities are computed by selecting
tracers in a pencil along the line of sight and estimating the
one-dimensional density distribution using optimal kernel smoothing
(Silverman 1986).  The quantity $\sigma_d$
denotes the half width corresponding to the mass enclosed within one
Gaussian standard deviation.  To convert to scale height $h$ of the
equivalent isothermal slab, an explicit evaluation determines that
$\sigma_d\approx1.8h$.  With a $45^\circ$ inclination, this is
$\sigma_d\approx2.6h$.  The variance of the velocity distribution of
stars along the line of sight is denoted $\sigma^2_v$.

Figure \ref{fig:thick} shows $h$ and $\sigma_v$.  There are two clear
trends: 1) the thickness of the disk increases; and 2) the velocity
dispersion very slowly {\em decreases}.  The slope, shown as a
straight solid line in Figure \ref{fig:thick}, is $70\pc/\gyr$.  The
evolved n-body disk has an approximately exponential profile, as also
predicted by the semi-analytic computation.  The linear increase in
$h$ with time is predicted by the underlying resonance theory and is
a natural consequence of secular evolution.

The decrease in velocity dispersion seems counterintuitive at first
glance but is often seen in globular cluster evolution.  In a fixed
gravitational potential, the heating would go into kinetic energy.
However, the work done on the {\em self-gravitating} disk decreases the
depth of the potential well and, by the virial theorem, also decreases
the kinetic energy.  The relative velocity dispersion tends to
increase owing to the increased orbital eccentricity and a larger
projection of the velocity along the line of sight caused by
increasing orbital inclination.  However this is not enough to offset
the overall decrease in kinetic energy and the magnitude of the
dispersion decreases.

In summary, the effect of the heating is significant although not as
dramatic as in the analytic computation that ignores the disk's self
gravity.

\begin{figure*}[th]
  \mbox{ \mbox{\epsfxsize=2.33in\epsfbox{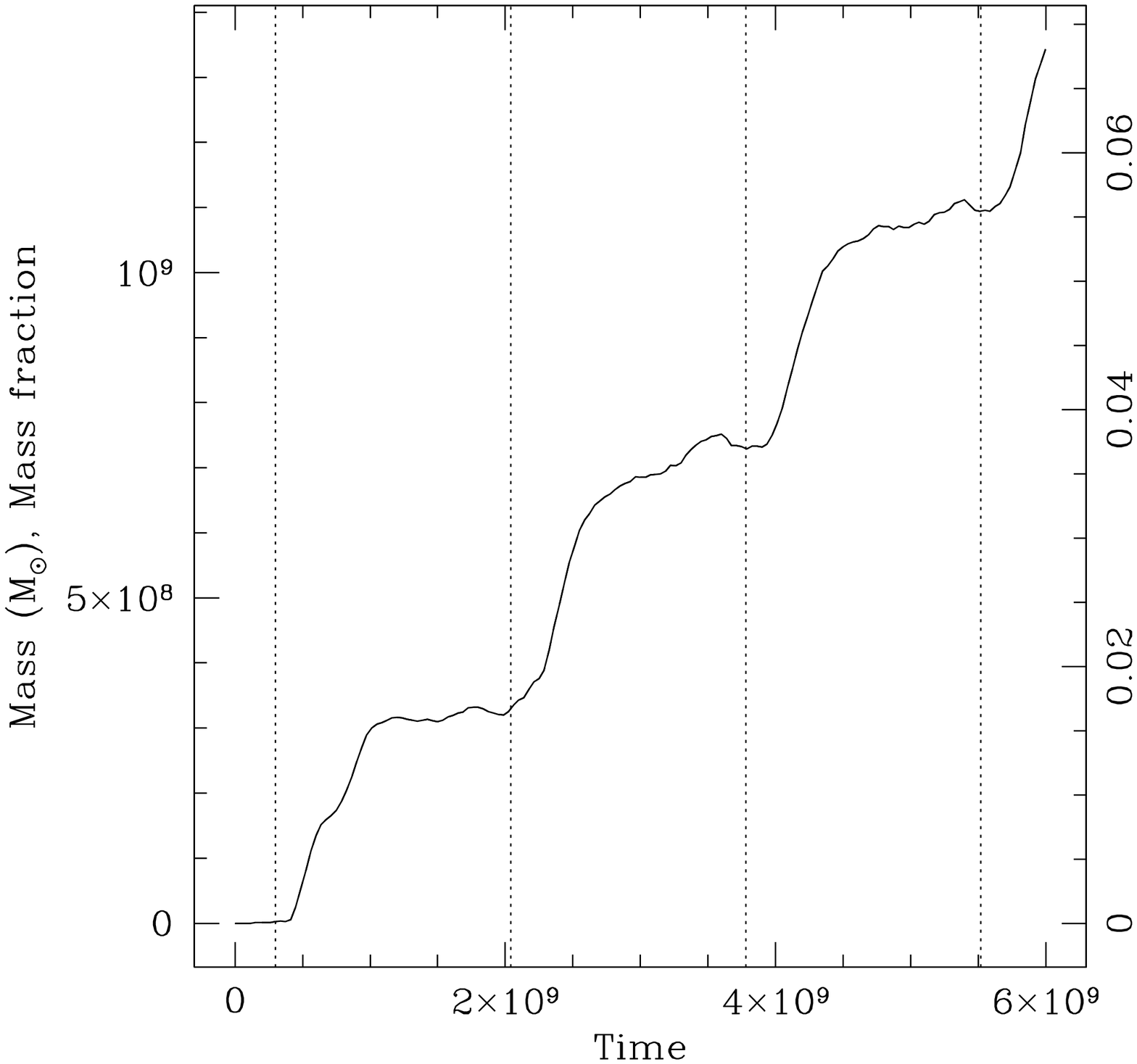}}
    \mbox{\epsfxsize=2.33in\epsfbox{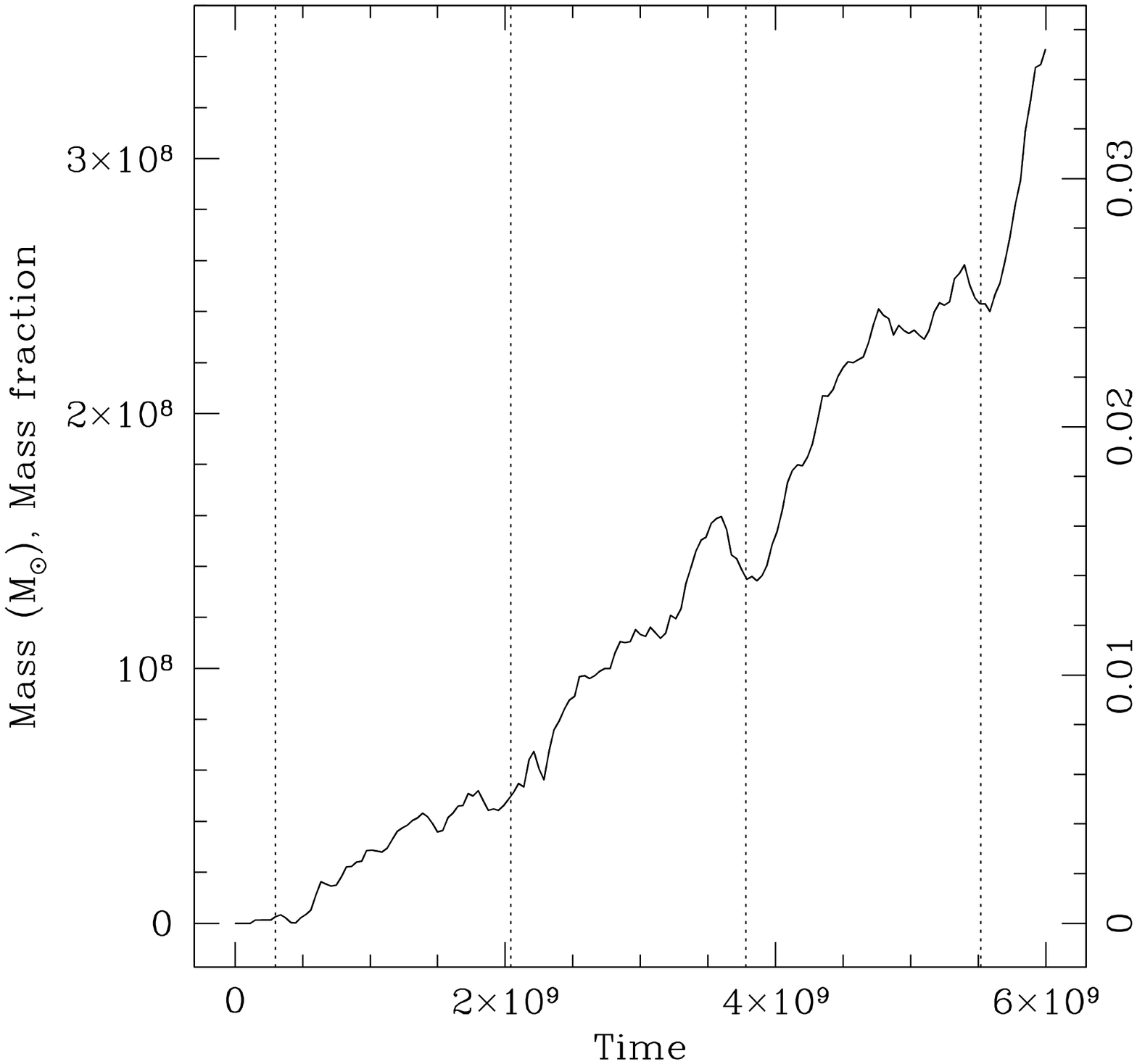}}
    \mbox{\epsfxsize=2.33in\epsfbox{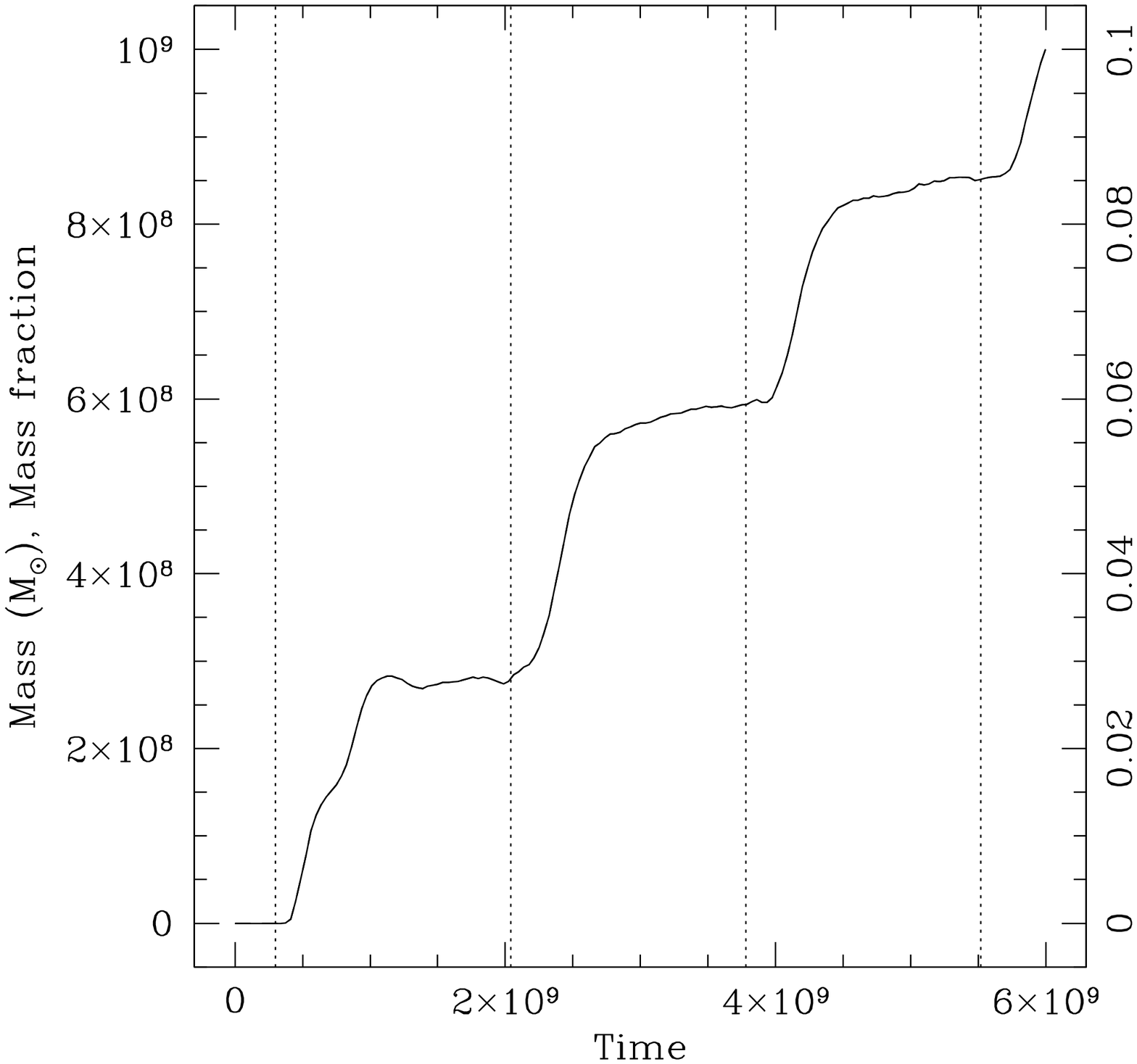}} }
  \caption{Mass loss as a function of time for total mass (left),
    LMC disk (center), and LMC halo (right).  In each panel, the left
    axis describes mass in solar masses and the right describes mass
    fraction.  The vertical dotted lines indicate perigalatica.
    \label{fig:massloss}
    }
\end{figure*}

\begin{figure*}[p]
  \mbox{ \mbox{\epsfxsize=3.5in\epsfbox{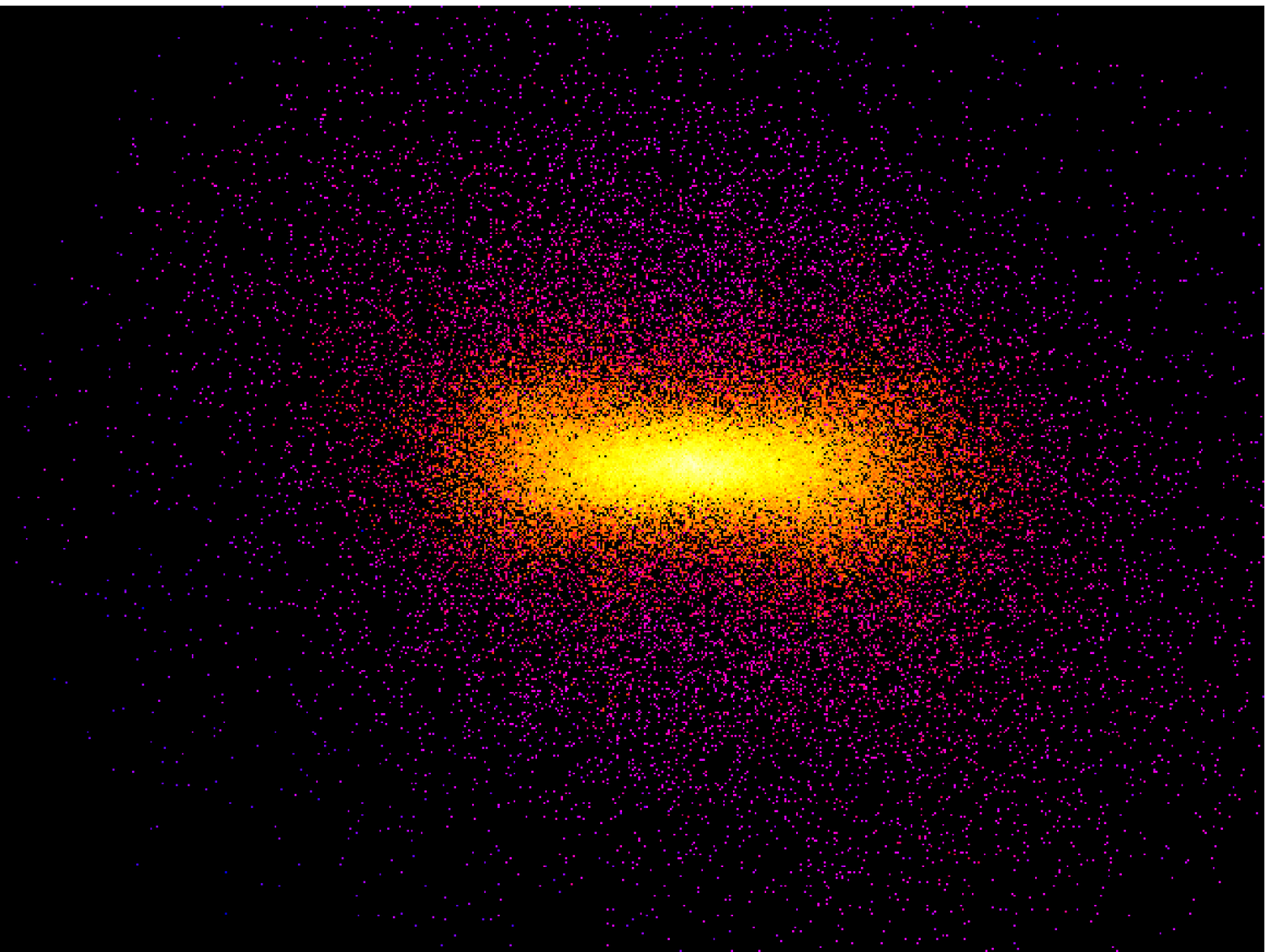}}
    \mbox{\epsfxsize=3.5in\epsfbox{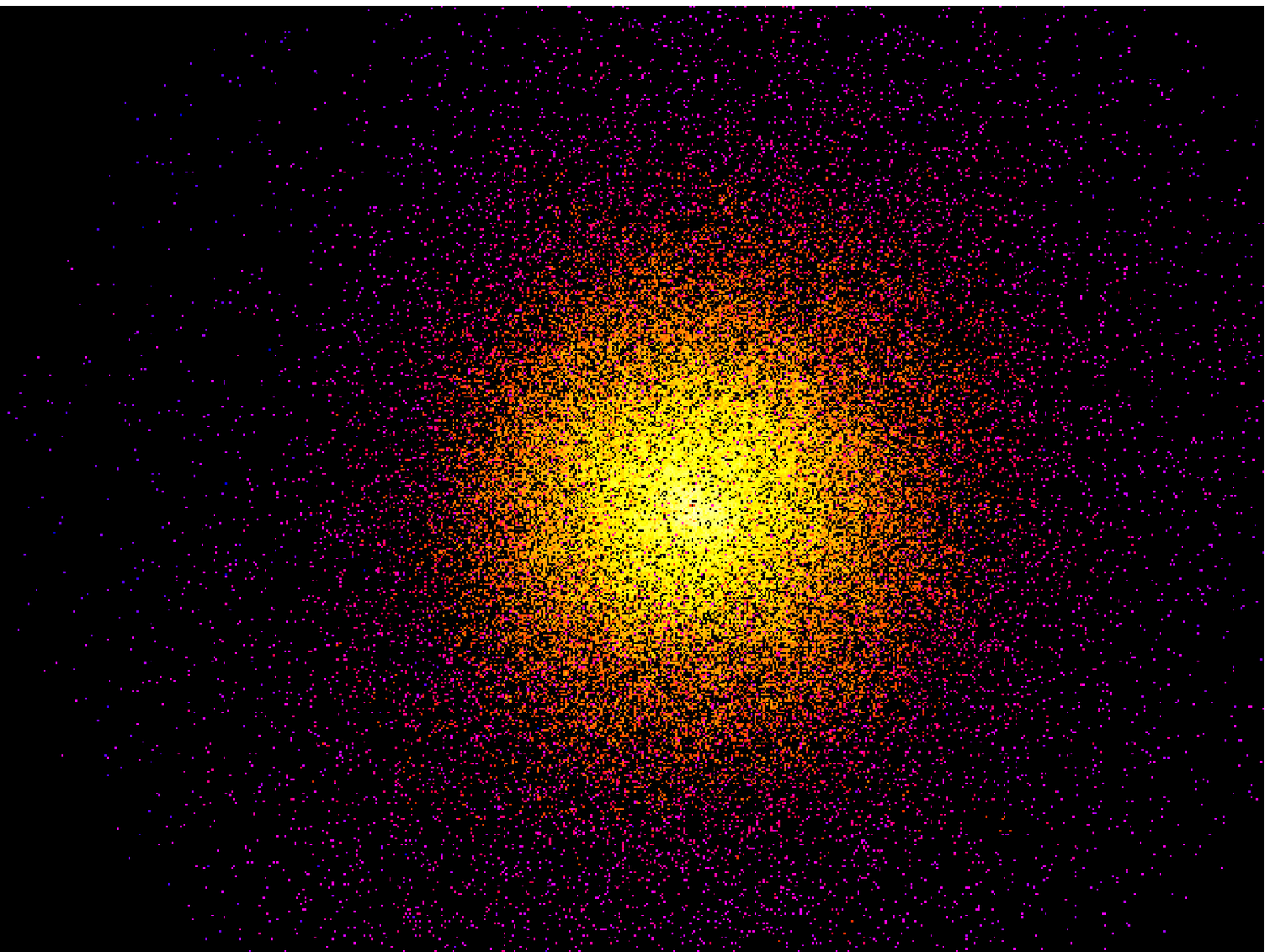}} }
  \caption{Edge-on (left) and face-on (right) views of the LMC disk
    after about 5 Gyr.  The points are color coded to indicate mass
    density on a logarithmic scale from blue to yellow.  The distance
    top to bottom is approximately $20\kpc$.
    \label{fig:nbody_close}
    } \vspace{20pt} \mbox{
    \mbox{\epsfxsize=3.5in\epsfbox{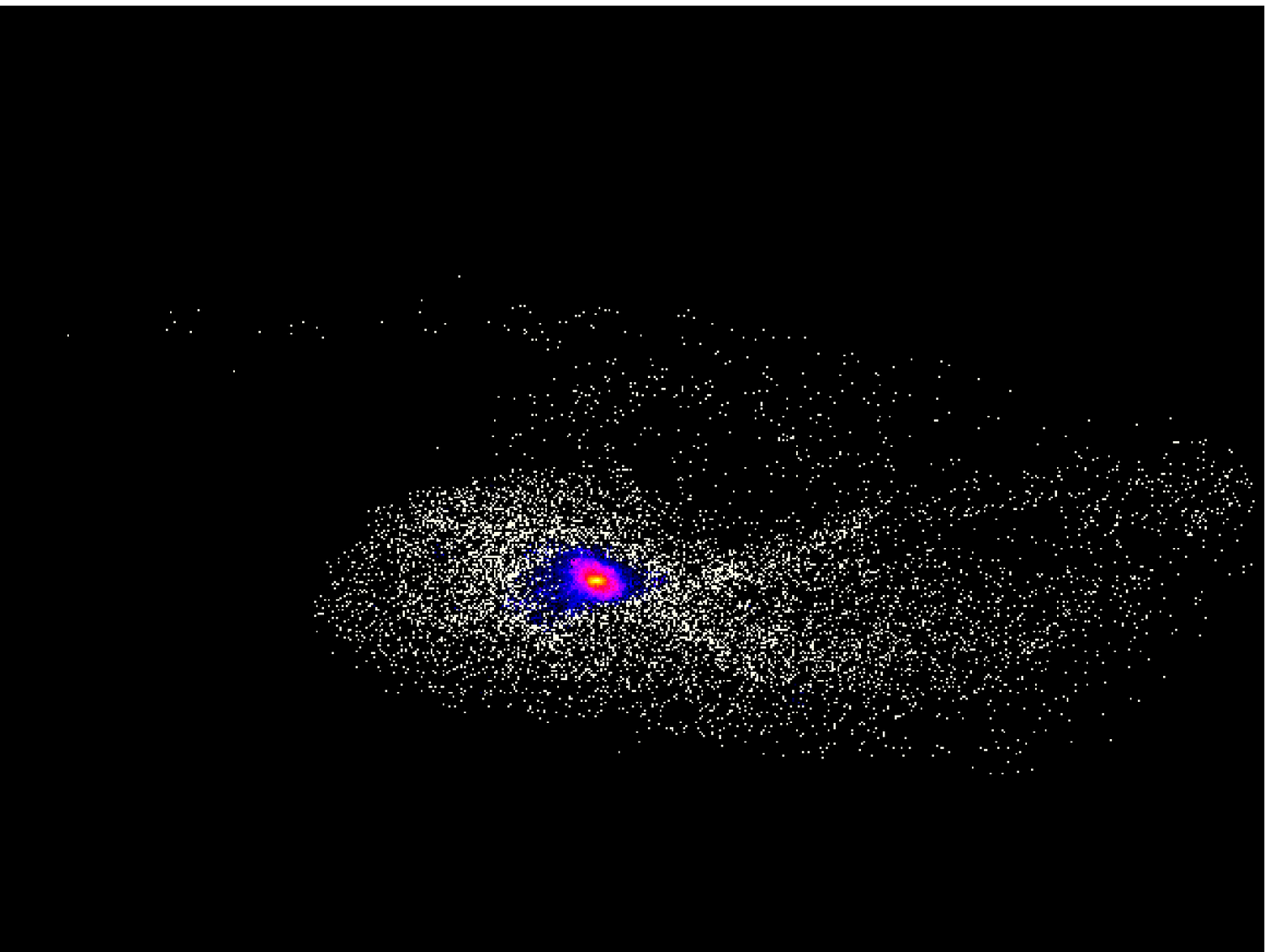}}
    \mbox{\epsfxsize=3.5in\epsfbox{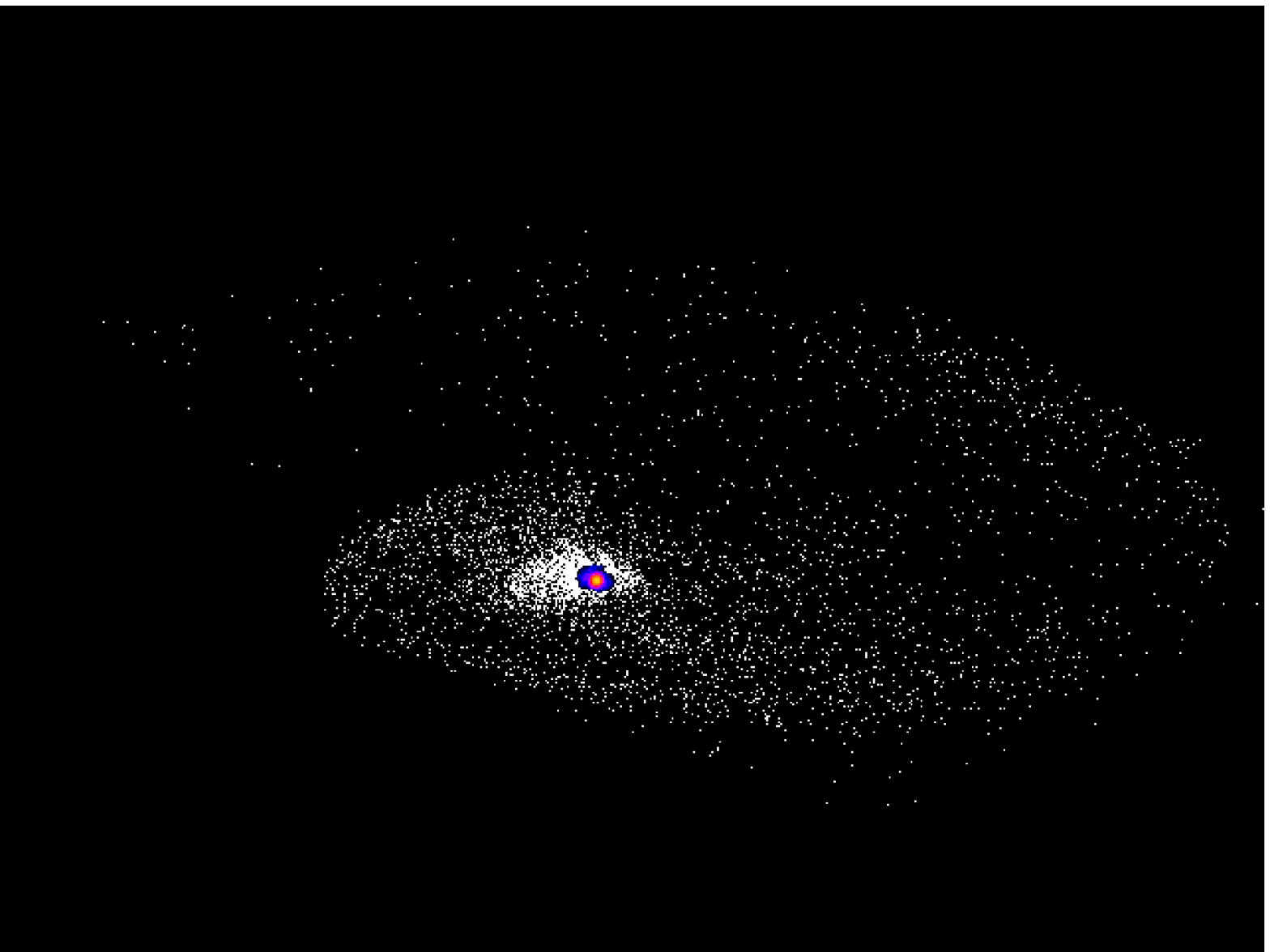}} }
  \caption{Large-scale views of evolved LMC, highlighting the
    low-density ejected material (white).  The sharp edge is caustic
    due to stars lost at a previous perigalacticon.  The distance top
    to bottom is approximately $50\kpc$.  The radius of the
    well-defined disk is approximately $10\kpc$.
    \label{fig:nbody_tails}
    }
\end{figure*}

Figure \ref{fig:massloss} describes the mass loss as a function of
time.  Mass beyond the LMC tidal radius is assumed to be lost.  Loss
of dark halo material, disk stars and total are shown separately with
the orbital pericenters indicated as vertical dotted lines.  Roughly
10\% of the halo and 3\% of the disk is lost by $T=6\gyr$.  The halo
material is lost episodically, with the peak loss just past every
pericenter.  The disk stars are lost at a roughly steady rate.  A
total mass of $1.4\times10^9\msun$ or 7\% of the original mass is lost
by $T=6\gyr$.  The spatial distribution of the LMC disk at this point
is shown in Figure \ref{fig:nbody_close} in both edge-on and face-on
projection.  The colors here range from blue to red, orange, yellow
and finally white as the mass density increases logarithmically.  The
outlying stellar spheroid that has been heated out of the disk is
fairly tenuous and the edge-on disk is thinner than it appears.
Figure \ref{fig:nbody_tails} highlights the distribution of lost mass
for both stars and halo material.  In this figure, the lowest
densities are shown as white.  Recall, the relative number of points
at different radii in these plots do not trace mass; the lower binding
energies are preferentially represented as described in
\S\ref{sec:params} in order to better resolve the mass loss.
  
\begin{figure*}[ht]
  \mbox{
    \vbox to 3.5in {
      \epsfxsize=3.0in\epsfbox{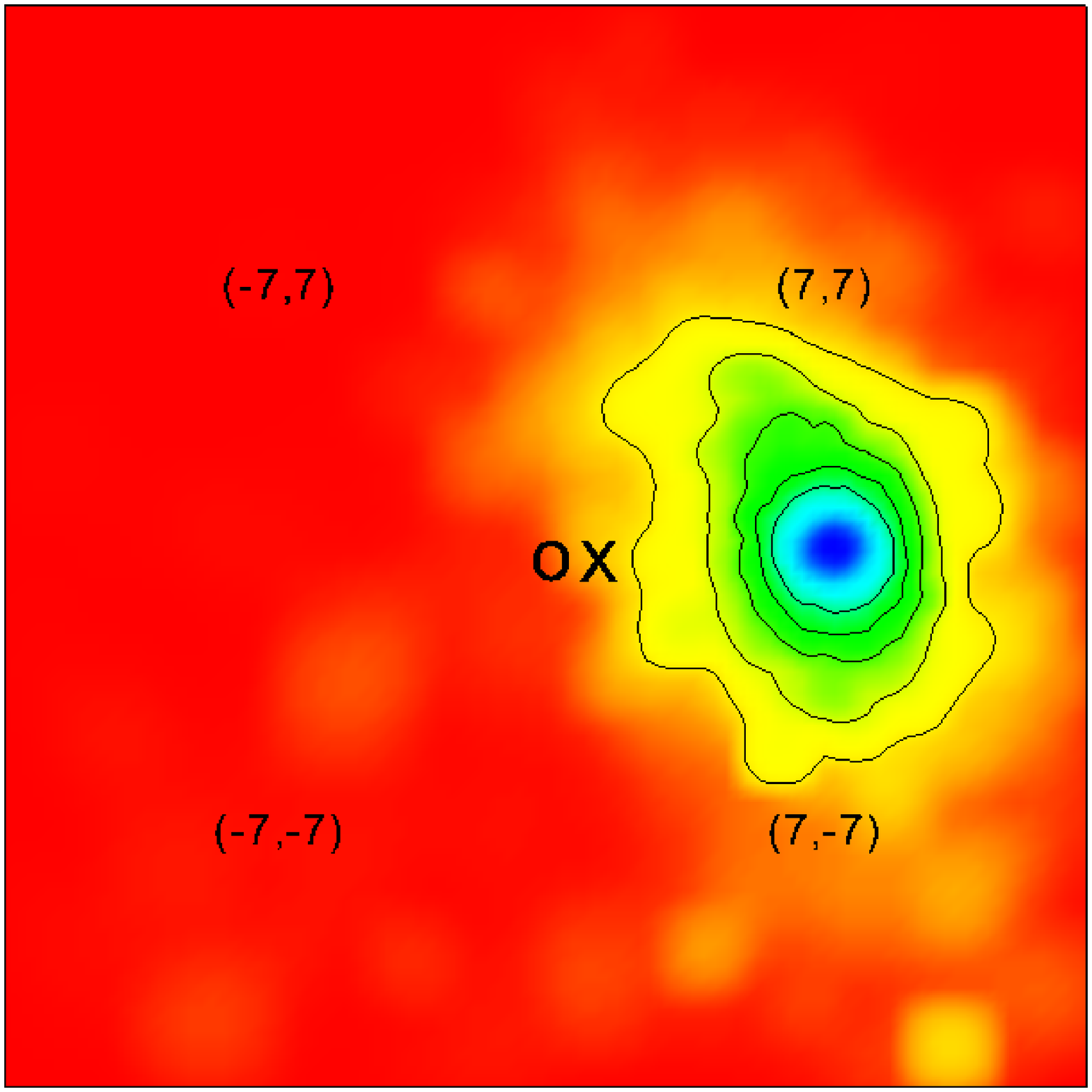}
      \hsize=3.0in
      \caption{
        The projected stellar density perpendicular to the orbital
        plane.  The Cloud is near pericenter and moving in the
        \protect{$\hat y$} direction. The five contours are spaced
        logarithmically and correspond to
        $1.5\times10^{-3}\msun/\pc^2$ to $1.5\times10^1\msun/\pc^2$.
        Each unit of length in the simulation is $7\kpc$; the points
        $(\pm7,\pm7)$ are labelled.  To provide a sense of scale, the
        `O' and `X' denote the Galactic center and an offset of
        $8\kpc$, respectively.
        \label{fig:cartxy190}
        }
      }
    \hspace{0.5in}
    \vbox to 3.5in {
      \vfill
      \epsfxsize=3.5in\epsfbox{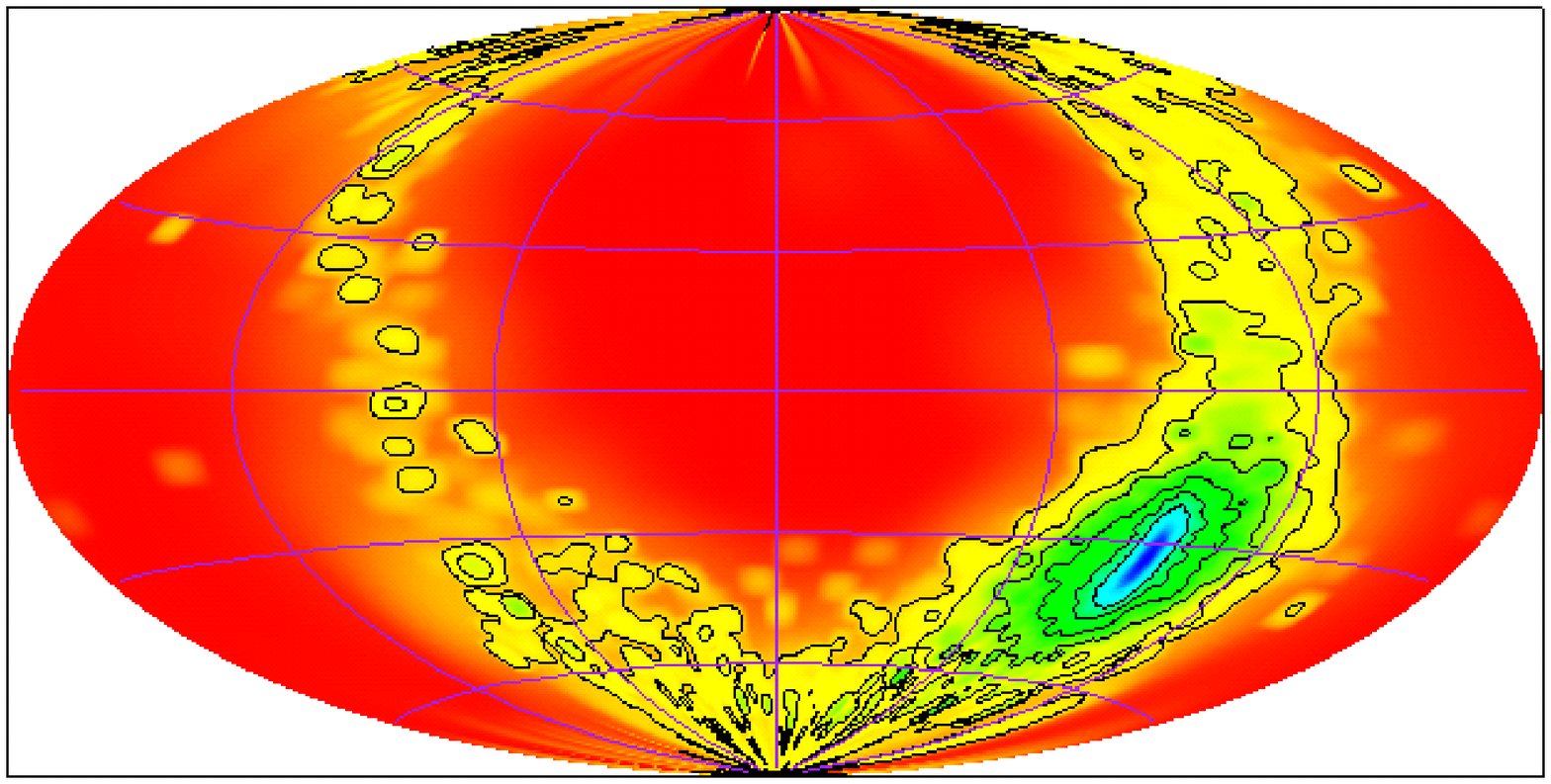}
      \vfill
      \hsize=3.5in
      \caption{
        The projected stellar density on the sky in Aitoff projection.
        The five contours are spaced logarithmically and correspond to
        $2\times10^1$ to $2\times10^5$ solar masses per square arc
        minute.  The simulation is shown with the LMC close to its
        current position.
        \label{fig:stream190}
        }
      }
    }
\end{figure*}

\subsection{Location of stripped mass}

There have been a number of searches for stars associated with the
observed gaseous Magellanic Stream (Sanduleak 1980, Recillas-Crus
1982, Brueck \& Hawkins 1983, Kunkel et al. 1997a, Guhathakurta \&
Reitzel 1998).  All but one of these studies target the
H{\small I} distribution.  Kunkel et al.  study carbon stars in the
outskirts of the LMC.

This paper suggests that the stellar debris has a different
distribution than the gaseous stream.  Tidal heating slowly advects
the orbits to both high inclination and radius before they are lost to
the Galactic potential.  The stripped distribution, then, has a larger
velocity dispersion than the kinematically cold disk component and is
not collimated on the sky.  This debris should not be expected to
track the stripped gas, which can dissipate and interact with the
Galactic halo gas.  A simulation that investigates this same scenario
but includes gas dynamics is in preparation.

After three orbits, the LMC is surrounded by stripped material (cf.
Fig. \ref{fig:cartxy190}).  Logarithmic contours and color coding
highlight the low-density distribution.  Note the low-level plateau of
material surrounding the bound LMC.  Figure \ref{fig:stream190} shows
the location of the predicted stream in Galactic coordinates with
logarithmic contour levels of projected mass density\footnote{Figures
  \protect{\ref{fig:cartxy190}} and \protect{\ref{fig:stream190}} were
  made by kernel smoothing the n-body distribution onto a rectangular
  grid.  In the latter case, the grid is first constructed in $l$ and
  $\cos b$ and rendered using an equal-area Hammer-Aitoff mapping.
  Unfortunately, the distortion due to the mapped bin shape is visible
  especially at the poles but the basic features are clear.}.  The
stellar ejecta would spread along a great circle across the sky if
viewed from the Galactic center.  The low star count density predicted
here, a few stars per square arc minute, is probably too low to have
been observed in small-area fields.  For example, the limit of
Guhathakurta \& Reitzel (1998), based on deep photometry centered on
the MS{\small IV} gas clump, is near but above the predicted projected
stellar mass density in debris stars.  However, the stripped material
along the LMC orbit but still outside the LMC should be detectable by
filtering a large-area survey along the predicted great circle.  Such
an analysis will be straightforward with a full-sky survey such as
2MASS.  The carbon stars observed by Kunkel et al. (1997) in the
outskirts of the LMC may be evidence for stellar debris near but
outside the Cloud.

Comparing both Figures \ref{fig:cartxy190} and \ref{fig:stream190}, it
is clear that the distribution of stripped stars on the sky are likely
to have a large spread in distance.  Of particular interest is the
extension in front of the LMC which may be a possible explanation for
the observed intervening stellar population toward the LMC reported by
Zaritsky \& Lin (1997).

\section{Microlensing}
\label{sec:mulens}

An extended LMC stellar distribution, both bound and unbound, can
enhance the microlensing optical depth caused by self-lensing.  We can
calculate the optical depth due to microlensing by using the estimated
density distribution from the n-body simulation (see Appendix for
details).  We use the same Galactic halo model adopted by the MACHO
collaboration for consistency (e.g. Alcock et al.
1997):
\begin{equation}
\rho_H = 0.0079 \: {R_0^2 + a^2 \over r^2 + a^2} \; M_\odot/pc^3,
\end{equation}
where $r$ is the Galactocentric radius, $R_0 = 8.5 \kpc$ is the
Galactocentric distance of the Sun and $a \approx 5\kpc$ is the core
radius.

The optical depth averaged along the line-of-sight is given by
\begin{equation}
\tau = \int _0 ^\infty \tau (D_s) p (D_s) dD_s \left[ \int _0 ^\infty
  p (D_s) dD_s \right]^{-1},
\end{equation}
where
\begin{equation}
\tau (D_s) = {4 \pi G \over c^2} \int _0 ^{D_s} \rho_d (D_d) {D_d (D_s
  - D_d) \over D_s} dD_d
\end{equation}
is the optical depth due to sources at a distance $D_s$, $\rho_d$ and
$\rho_s$ are the lens and source densities respectively, and
\begin{equation}
p (D_s) dD_s = C \rho_s (D_s) D_s^{2+2\beta} dD_s
\end{equation}
(Kiraga \& Paczy\'nski 1994) is the
probability of finding a source in the interval $\left[D_s,
  D_s+dD_s\right]$.  We take $\beta = -1$, consistent with a fit to
the Bahcall-Soneira model (1980).

\begin{figure*}[th] 
  \mbox{ 
    \mbox{\epsfxsize=3.5in\epsfbox{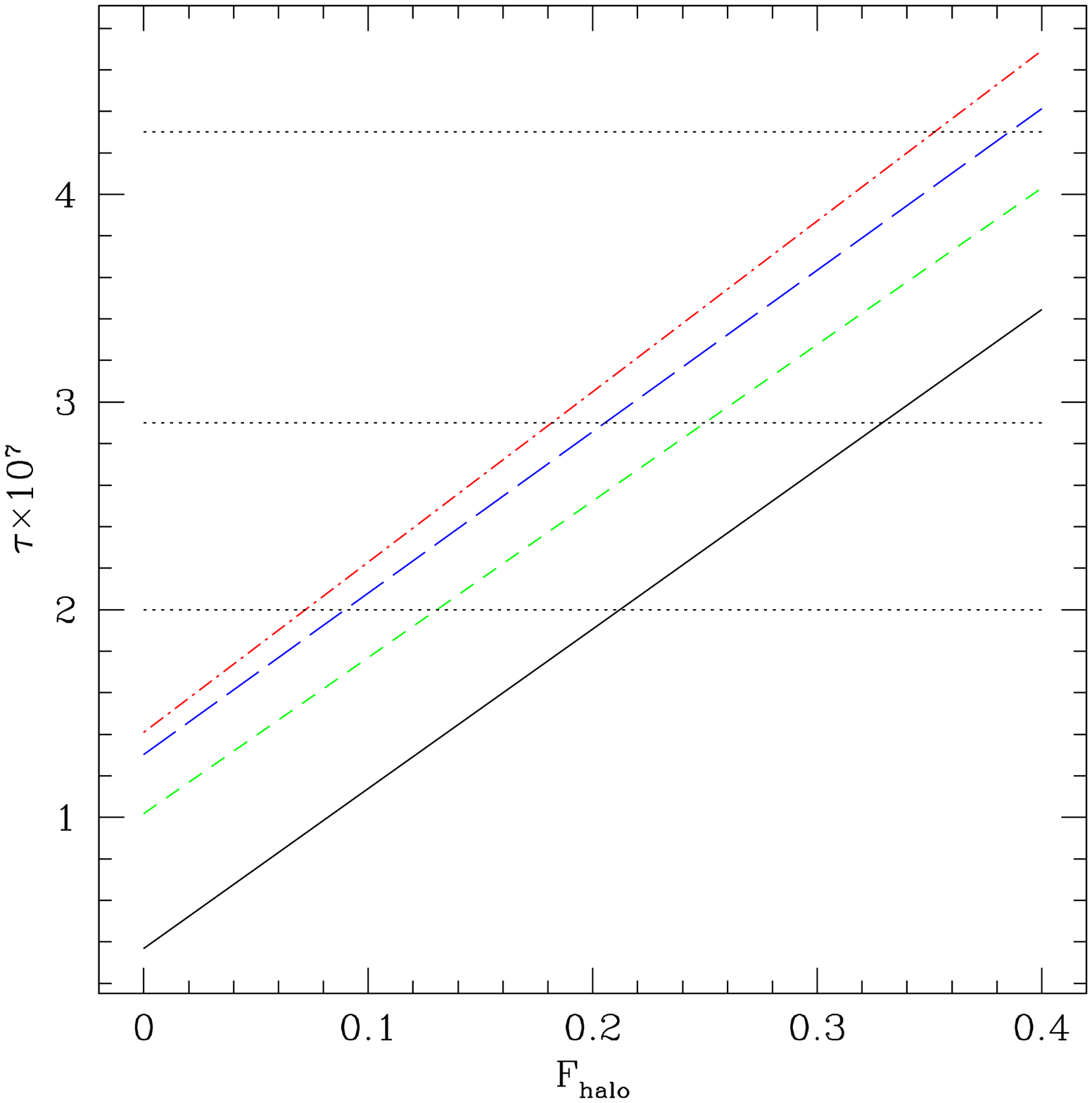}}
    \mbox{\epsfxsize=3.5in\epsfbox{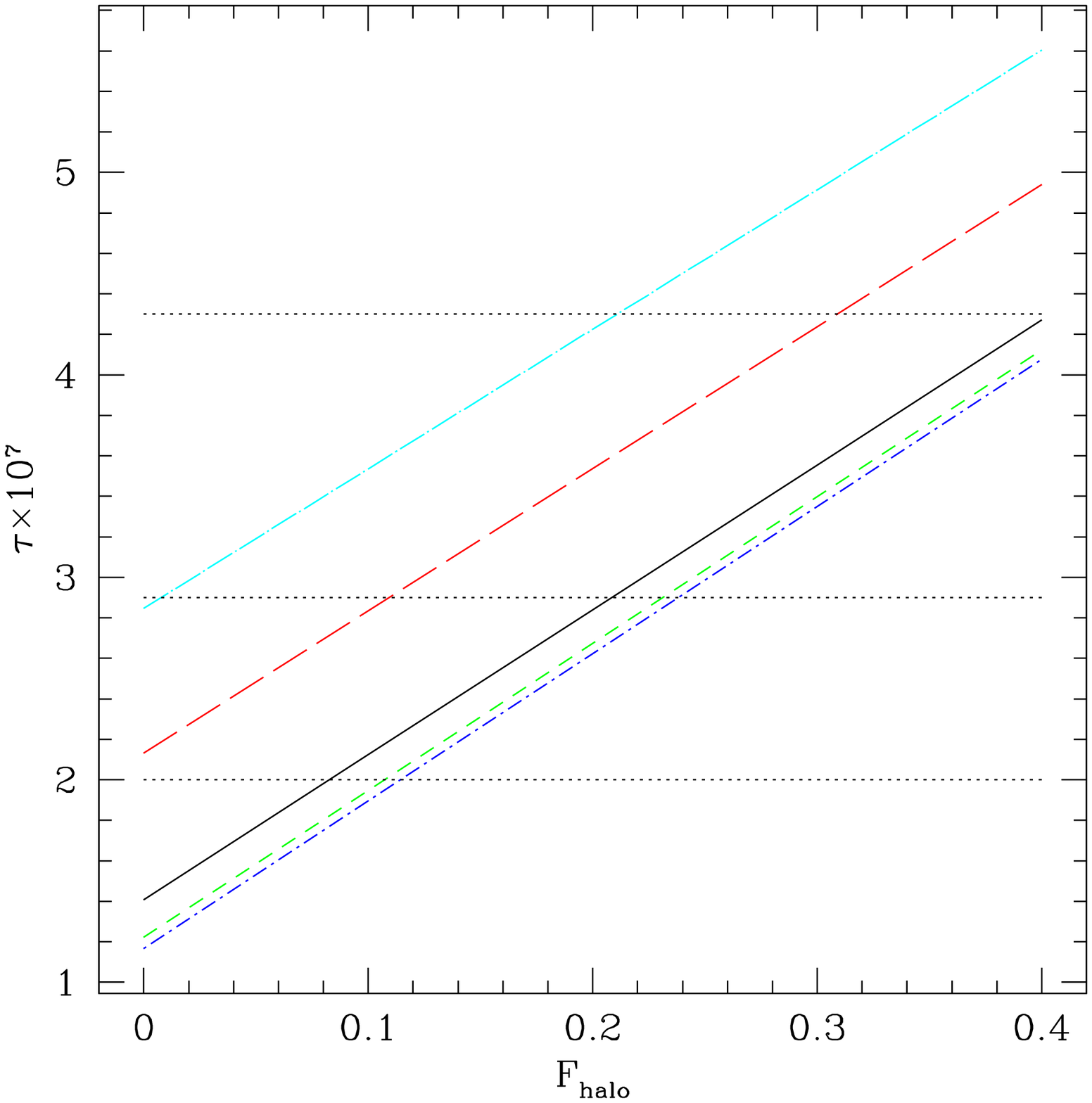}}
    }
  \caption{Left: microlensing optical depth as a function of MACHO
    fraction in the Galactic halo including MACHOs in LMC halo.  The
    middle (upper, lower) horizontal dotted line show the observed
    microlensing ($\pm 1\sigma$ confidence limits) from Alcock et al.
    (1997).  Depth computed using the Kiraga \& Pacsy\'nzki $\beta$
    parameterization with $\beta = -1$.  The four curves shows the
    predicted microlensing of the initial state (solid) and three
    successive pericenters (short-dash, long-dash, and dash-dot,
    respectively).  Right: shows variation of microlensing optical
    depth as a function of disk inclination for the final pericenter
    shown at the left.  The five curves show an inclination of 11.25,
    22.5, 45, 67.5, and 78.25 degrees from bottom to top (short dash
    dot, short dash, solid, long dash, long dash dot, respectively).
    \label{fig:optdepth}
    }
\end{figure*}

For these simulation-based estimates, the LMC location is chosen at a
point in its orbit that matches its present position.  Unfortunately,
this does not guarantee that the orientation of the disk in the
simulation corresponds to the one observed.  Rather than perform
expensive iterations, the coordinates are transformed to the observed
true orientation.  The line-of-sight density distribution is computed
using the kernel smoothing procedure described in \S\ref{sec:results}.

First we assume no Galactic halo MACHOs; both source density $\rho_s$
and deflector density $\rho_d$ include only the stellar LMC
distribution.  This gives a total optical depth due to LMC
self-lensing of $1.4 \times 10^{-7}$ at the end of three LMC orbits
(5.5 Gyr) in the simulation.  This falls shy of the observed value,
$2.9^{+1.4} _{-0.9} \times 10^{-7}$, by nearly two standard deviations
although precise comparison is impossible since the simulation does
not follow the entire LMC history.  Nonetheless, self-lensing
including the tidally evolved distribution is a significant
contribution to the optical depth.  The best fit value is
$F_{halo}\approx0.21$ for the final orbit.  If the Milky Way halo
contains MACHOs, it is likely that the LMC halo also contains the same
fraction.  The LMC halo has one half of the total mass initially.  In
this case, the best fit is obtained for $F_{halo}\approx 0.18$.
Figure \ref{fig:optdepth} (left) shows the run of $\tau$ with
$F_{halo}$ for this latter case.

The increase in the contribution to microlensing optical depth is
dominated by the thickened disk rather than the lost stars in this
simulation.  Although mass is being lost continuously, the density
profile near the disk is slowly changing after the first few orbits
(as in Fig. \ref{fig:lmcdisk1} and reflected in Fig.
\ref{fig:optdepth} (left) for $F_{halo}=0$).  However, this makes the
self-lensing a strong function of disk inclination as shown in Figure
\ref{fig:optdepth} (right).  For example, an inclination of $67.5$,
$78.25$ degrees would imply $F_{halo}=0.11, 0.0$, respectively.  This
is sensitivity is one-sided; decreasing the inclination below 45
degrees make little change in the $\tau$ estimates.

In summary, the tidal disk heating makes a significant contribution to
self-lensing.  For no MACHOs, $F_{halo}=0$, the optical depth of the
tidally evolved disk is three times larger than the initial $\sech^2$
disk.  This translates to a factor of two difference in the best
estimate of $F_{halo}$ (cf. Fig \ref{fig:lmcdisk1}, left) and
decreases the significance of rejecting the $F_{halo}=0$ hypothesis.

\section{Discussion}
\label{sec:discussion}

Although the physics of resonant heating and secular evolution applied
here to thicken the disk and augment tidal stripping is
well-understood and secure, several sources of wiggle room remain.
First, the n-body simulations are technically difficult.  There is no
analytic method for constructing an equilibrium in a time-dependent
tidal field so the initial model must come to a new equilibrium to
start.  The new virialized equilibrium has a weak rotating $m=1$
distortion that no doubt increases the heating rate although the disk
heating during this initial period appears to be minimal.  Because the
simulation has been followed for nearly 5 Gyr and therefore many
dynamical times, we took great care to estimate the rate of secular
evolution due to intrinsic fluctuations.  Both $m=1$ mode and
fluctuation heating are smaller than the tidal effects.  However, the
heating from global excitation by noise or other sources can produce
thickening and needs to be understood and treated with care by
simulators.

A second uncertainty is the unknown initial conditions for the LMC.
The strongest evidence for the existence of some dark component is the
weakly falling rotation curve indicated by globular clusters and
planetary nebula (e.g.  Schommer et al. 1992).  This led to adopting
an even mass split between the disk and halo components.  Given the
relatively short time for scale height growth why does the LMC appear
to have a well-defined disk?  A much more massive and an extended LMC
halo would protect the luminous distribution from tidal stripping.  On
the other hand, such a halo is limited by self-lensing (assuming that
it contains the same MACHOs attributable to the Galactic halo),
constrained by the LMC rotation curve, would decrease the orbital
decay time, and may be untenable given the observed SMC kinematics.
In addition, such a halo is more readily stripped than the disk and
the work done on the disk by the readjusting gravitational potential
promotes heating in addition to the halo-disk coupling mentioned
above.

Stripping is a natural consequence of the LMC--Milky Way interaction.
A definitive failure to detect a stripped stellar component will
necessitate a reevaluation of LMC structure.  A speculative
possibility is that a more massive LMC recently lost equilibrium in
the Milky Way tidal field.  The exposure of the disk to a significant
tidal force might be recent.  In such a scenario, the SMC most likely
was a satellite of the larger primordial LMC and is now interacting
directly with its luminous gas-rich disk.  More generally, these
dynamical mechanisms will affect all Magellanic-like systems and may
help constrain their histories and determine the extent of their dark
matter halos.

The current interest in MACHO detections and limits motivated some
simple estimates of self-lensing by the LMC or by tidally-stripped LMC
stars.  Because the LMC orbit is fixed in the treatment here.  the
trail of unbound material at the end of the simulation is probably
less extended than one might expect in Nature.  Nonetheless, the
self-lensing is dominated by material in the outer parts of the LMC or
recently unbound and therefore this idealization seems unlikely to be
significant.  The simulation suggests an extension in front of the LMC
due to material lost at or near pericenter that may be a possible
explanation for the observed intervening stellar population
toward the LMC reported by Zaritsky \& Lin
(1997).

\section{Summary}
\label{sec:summary}

The major conclusions of this paper are as follows:

\begin{enumerate}
\item {\em The Milky Way is a significant evolutionary driver of LMC
    structure.}  The time-dependent tidal forcing by the Milky Way
  will heat the LMC disk, producing extended rotating spheroid
  component.
\item {\em We find that the disk scale height increases at a rate of
    $70\pc/\gyr$ (cf. Fig \ref{fig:thick}).}  The heating has several
  components.  First, there is a direct resonant coupling between the
  time dependence of tidal forcing and the stellar orbits within the
  LMC disk.  Second, the body torque from the Milky Way causes the LMC
  disk to precess.  The interaction between the LMC halo and its
  precessing disk heats the disk.  This is a {\em new} but important
  mechanism for heating the disks of satellites.
\item {\em The stellar velocity dispersion decreases due to disk
    heating.}  The work done against the LMC gravitational potential
  decreases the depth of the potential well and the new
  quasi-equilibrium, although more extended, requires less velocity
  support. The sign of the effect follows from the virial theorem.
  Although a surprise to some, this effect has been well-documented
  for the evolution of star clusters.
\item {\em The mass loss rate is approximately $3\times10^8\msun$ per
    orbit or roughly 2\% per orbit at the current time.}  The fraction
  of halo loss to disk loss is roughly 3:1.
\item Because the heated, extended component is preferentially lost to
  tidal stripping, {\em the unbound stars will not be distributed like
    the Magellanic gas stream but in a diffuse distribution about the
    LMC.}  This component may be a source of both microlensing sources
  and lenses and affect MACHO estimates.  Overall, we estimate that
  the heated disk and tidally stripped component may make a
  significant contribution to gravitational microlensing.
\end{enumerate}

\acknowledgments I thank Neal Katz and Sergei Nikolaev for many useful
discussions and Neal Katz and Eric Linder for comments on the
manuscript.  This work described here was supported in part by NSF
AST-9529328 and NASA/JPL 961055.

\appendix
\section{LMC parameters}

\subsection{The LMC tidal radius and mass}

Star count maps of the outer LMC (e.g. Irwin 1991)
show an extended distribution with a fairly sharp edge, typical of a
tidally truncated system.  To get an independent measurement using
2MASS star counts, we selected 12 subfields $0.5^\circ\times0.5^\circ$
in size which probe the LMC halo at the projected radii of $2^\circ -
5^\circ$ from the LMC center ($l_{II}=280.5^\circ,
b_{II}=-32.9^\circ$).  The counts were fit to Gaussian and power-law
spherical models, $\rho \propto e^{-{r^2 /2 a^2}}$ and $\rho \propto
\left( 1 + r^2/a^2 \right)^{-\gamma}$, using a maximum likelihood
procedure.  The simple analytic forms for these profiles make the
likelihood computation feasible.  To estimate the mass of the LMC, we
fit these analytic profiles by King models to estimate the tidal
radius:
\begin{equation}
M_{LMC} = \left( r_t \over R_{LMC} \right)^3 2 M_{MW},
\end{equation}
where $R_{LMC}$ is the distance to the LMC and $M_{MW} = 5 \times
10^{11}\;M_\odot$ is the mass of the Milky Way.  This is a total mass
estimate, including both the halo and the disk mass.  

This procedure will underestimate the mass for two reasons.  First,
simulations suggest that the observed $r_t$ is 75\%--80\% of the
dynamical critical point.  Second, a tidally-limited object is likely
to be elongated toward the Galactic center and therefore roughly along
the line of sight.  For a centrally-concentrated object, the axis
ratio is $a/c=1.5$.  The first correction yields a factor of
$(10/8)^3\approx 2$.  The second increases the enclosed volume by
roughly $3/2$ but whether or not this should be included depends on
orientation.  A reasonable correction factor is then between 2 and 3
and we conservatively choose the former.  The parameters of the `best
fit' models are $a=2.6, 2.8$ for the Gaussian and power-law model with
$\gamma=2$, respectively.  For both cases, the lower mass limit is
$1\times10^{10}\msun$ with a best estimate of $2\times10^{10}\msun$.
An in-depth presentation of these results is in preparation.

As an independent check, we made a naive estimate of the mass of the
LMC from the analysis of the halo population using the star counts in
our fields.  Most of the sources observed by 2MASS are M-giants with
the absolute magnitude in $K$-band $K < -4^m$ (for the distance to the
LMC of $50\,\kpc$ and 2MASS $K_s$-band $SNR=10$ flux limit of
$14.3^m$).  Assuming that these M giants are representative of an
intermediate age population with the extended distribution derived
above, we may estimate the total stellar mass using an infrared
luminosity function.  For this purpose, we adopt the Galactic
luminosity function in Wainscoat et al.
(1992).  Integrating over the
luminosity function with a standard luminosity-mass relation results
in stellar mass of $\approx 4 \times 10^9\;M_\odot$, which is
consistent with these estimates.

\subsection{Rotation curves: a consistency check}

Schommer et al. (1992) summarizes the derived rotation curve for
clusters, planetary nebulae and H{\sc I} including the Meatheringham
et al. results (see Schommer et al. Fig. 8).  Using a luminosity
function derived for an exponential disk and typical velocity
dispersion in halo with a flat rotation curve, one finds that the
circular velocity $V_c$ is roughly 20--30\% larger than the rotation
value $V_o$. For a rotation curve with $V_o\approx75$ km/s, one finds
a mass within 10.8 kpc of $M\gta2\times10^{10}\msun$ which is nicely
consistent with the tidal radius estimate.  See Schommer et al. for
more extensive discussion of these arguments.

\section{Computational notes}

\subsection{Grid-based Boltzmann solution}

The evolution of a perturbed equilibrium can be explored with a
time-dependent perturbation theory (e.g. Weinberg 1994b, Murali \&
Weinberg 1997a).  The
physics behind this approach is as follows.  The period of the LMC
orbit is longer than the periods of many of the stellar orbits within
the cloud and such orbits are adiabatically invariant to the
time-dependent tidal forcing.  However, in cases where the frequencies
of the stellar orbit are commensurate with the forcing frequencies,
the resulting degeneracy breaks the adiabatic invariant.  The change
in the gravitational potential causes the resonance to sweep through
phase space as described in \S\ref{sec:boltz} and the direction of the
passage determines the effect of the resonance on the conserved
quantities.  The net effect on the gravitational potential depends on
the scale of the inhomogeneity in the phase-space distribution.  The
evolution equations, therefore, take the form of a collisional
Boltzmann equation with the right hand side depending on a gradient of
phase space.  Numerically, we approximate the solution of this
equation by a two step process:
\begin{enumerate}
\item Update the phase-space distribution function in a fixed
  potential using the finite-difference representation for the
  evolution term on the right hand side.  This term can be written in
  flux form as described below and therefore conserves density over
  one step.
\item Hold the distribution function fixed as a function of actions
  and solve for a new equilibrium.  For ease of solution, the
  potential is assumed to have spherical symmetry while the
  distribution itself may be generally axisymmetric (e.g. $f=f(E, L,
  L_z)$).
\end{enumerate}

\subsection{Resonant heating rates}

\def\pdrv#1#2{{\partial #1 \over \partial #2}}
\def\ldw{\bf l \cdot w}
\def\lv {\bf l}
\def\Iv {\bf I}
\def\wv {\bf w}
\def\ldomega {\bf l \cdot \Omega}
\def\ldw {\bf l \cdot w}
\def\ldf {{\bf l} \cdot\pdrv{f_0}{\bf I}}
\def\Omegav {\bf\Omega}
\def\AAsum {\sum_{\lv=-\infty}^{\infty}}
\def\Fsum {\sum_{n=-\infty}^{\infty}}
\def\Ebar {\langle\langle \dot E \rangle\rangle}

The linearized Boltzmann equation is a linear partial differential
equation in seven variables.  Using action-angle variables, we can
separate the equation and employ standard distribution functions
constructed according to Jeans' theorem (Binney \& Tremaine
1987).  The explicit form of the linearized
Boltzmann equation is
\begin{equation}
 \pdrv{f_1}{t}+\pdrv{f_1}{\wv}\pdrv{H_0}{\Iv}-
                \pdrv{f_0}{\Iv}\pdrv{H_1}{\wv}=0, 
\end{equation}

\noindent where $\wv$ is the vector of angles, and $\Iv$ are the conjugate 
actions.  The quantities $f_0$ and $H_0$ depend on the actions alone.
Making the assumption that the tidal force from the Galaxy is small,
the perturbation may be separated into phase-space and time
components, $H_1=\eta({\bf r})g(t)$, expanded in a Fourier series in
action-angle variables (e.g. Tremaine \& Weinberg
1984).  Each term $f_{1\lv}$ in the
Fourier series is the solution of the following differential equation:
\begin{equation}
  \pdrv{f_{1\lv}}{t}+(i\ldomega) f_{1\lv}=i\ldf V_{\lv}(\Iv)g(t)
  \equiv i\ldf H_{1\lv},
  \label{eq:lboltz}
\end{equation}
where $\Omegav=\partial H_0/\partial{\Iv}$ and
\begin{equation}
  V_{\lv}(\Iv)={1\over (2\pi)^3}\int_{-\pi}^{\pi}\eta({\bf r})e^{-i\ldw}d^3\wv.
\end{equation}
The quantity ${\bf l}$ is a vector of integers whose rank is the
number of degrees of freedom; e.g. for the three dimensional problems
considered here, ${\bf l}=(l_1,l_2,l_3)$.  In practice, we usually
confine our perturbation to a particular set of spherical harmonics or
cylindrical harmonics which restricts two out of three to a finite set
(see Tremaine \& Weinberg 1984 for details).

The rate of change in energy or action arising from the perturbation
follows from Hamilton's equations and is
\begin{eqnarray}
  {\dot E} &=& \AAsum i\ldomega H_{1-\lv}f_{\lv} \nonumber \\
  {\dot I}_j &=& \AAsum iI_j H_{1-\lv}f_{\lv}.
  \label{eq:eirates}
\end{eqnarray}
For periodic perturbations, the time dependent amplitude may be
represented by a Fourier series,
\begin{equation}
 g(t)=\Fsum a_n e^{i n\omega t}.
 \label{eq:fourexp}
\end{equation}
With this form for $g(t)$, equation (\ref{eq:lboltz}) may be solved
for the perturbed distribution function by Laplace transform.
Finally, phase averaging the quantities in equations (\ref{eq:eirates}
yields the following time-asymptotic rates:
\begin{eqnarray}
 \langle \dot E \rangle &=& -8\pi^4\AAsum(\ldomega)(\ldf) |V_{\lv}|^2 
        \Fsum |a_n|^2\delta(n\omega-\ldomega).  \label{eq:idot} \\
 \langle \dot I_j \rangle &=& -8\pi^4\AAsum l_j(\ldf) |V_{\lv}|^2 
        \Fsum |a_n|^2\delta(n\omega-\ldomega). \label{eq:edot}
\label{eq:res}
\end{eqnarray}
Murali \& Weinberg (1997a) show that this expansion, continued to the
next order, results in an equation for the change in distribution
function in terms of these rates which takes the form:
\begin{equation}
  \langle {\dot f}_2\rangle \propto
  {\partial\over\partial{\bf I}} \cdot \langle{\bf{\dot I}}\rangle
\end{equation}
which may be solved by standard flux-conserving finite-difference
methods.

\end{document}